\documentclass[a4paper,11pt,english]{article}

\usepackage[a4paper,left=3cm,right=3cm,bottom=3.5cm,top=3cm]{geometry}
\usepackage[textwidth=2.5cm]{todonotes}

\usepackage[T1]{fontenc} 
\usepackage{lmodern} 
\usepackage[utf8]{inputenc} 
\usepackage[ngerman, main=english]{babel} 
\usepackage[square]{natbib}
\usepackage{xcolor}
\usepackage{xspace}
\usepackage{verbatim}
\usepackage{float}
\usepackage{amsbsy}
\usepackage{marvosym}
\usepackage{mathtools}
\usepackage{calc}
\usepackage[framemethod=tikz]{mdframed}
\usepackage{lipsum}
\usepackage{placeins}
\usepackage{expl3}
\ExplSyntaxOn
\cs_new:Npn \displayasbinary #1 {\int_to_bin:n{#1}}
\ExplSyntaxOff
\usepackage{stringstrings}
\usepackage{bm}
\usepackage{fbox}
\usepackage{pgfplots}
\usepackage{setspace}
\usepackage{listings}
\usepackage{numprint} 
\usepackage{siunitx}
\usepackage{printlen}
\usepackage{relsize}
\usepackage{needspace}
\usepackage{thmtools}
\usepackage{thm-restate}
\usepackage{textcomp}
\usepackage{graphicx}
\usepackage{longtable}
\usepackage{rotating}
\usepackage{advdate}
\usepackage{lipsum}

\usepackage{amsmath}  
\usepackage{amssymb}  
\usepackage{stmaryrd} 
\usepackage{dsfont}   

\usepackage[margin=0pt,labelfont=bf]{caption} 

\usepackage{cite} 
\usepackage[babel,autostyle]{csquotes} 
\MakeOuterQuote{"}	
\usepackage{subcaption} 
\usepackage{enumerate} 
\usepackage{wrapfig} 
\usepackage{lscape}
\usepackage{rotating}
\usepackage{float} 
\usepackage{framed} 
\usepackage[framed,hyperref,amsmath,thmmarks]{ntheorem} 
\usepackage[unicode,pdfmenubar,linktoc=all,hidelinks,bookmarks]{hyperref} 
\usepackage[capitalise,nameinlink,noabbrev]{cleveref}
\usepackage{url} 
\usepackage{multicol} 

\usepackage{xspace}

\usepackage{eurosym}

\usepackage{pdflscape} 
\usepackage[hang]{footmisc}   
\usepackage{bibgerm}

\usepackage{makeidx} 
\makeindex

\usepackage[final]{listofsymbols}  

\definecolor{my-pink}{RGB}{233,30,99}
\definecolor{my-purple}{RGB}{156,39,176}
\definecolor{my-deep-puple}{RGB}{103,58,183}
\definecolor{my-indigo}{RGB}{63,81,181}
\definecolor{my-blue}{RGB}{33,150,243}
\definecolor{my-light-blue}{RGB}{3,169,244}
\definecolor{my-cyan}{RGB}{0,188,212}
\definecolor{my-teal}{RGB}{0,150,136}
\definecolor{my-green}{RGB}{76,175,80}
\definecolor{my-light-green}{RGB}{139,195,74}
\definecolor{my-lime}{RGB}{205,220,57}
\definecolor{my-yellow}{RGB}{255,235,59}
\definecolor{my-amber}{RGB}{255,193,7}
\definecolor{my-orange}{RGB}{255,152,0}
\definecolor{my-deep-orange}{RGB}{255,87,34}
\definecolor{my-brown}{RGB}{121,85,72}
\definecolor{my-grey}{RGB}{158,158,158}
\definecolor{my-blue-grey}{RGB}{96,125,139}
\colorlet{my-light-blue-grey}{my-blue-grey!40!white}

\colorlet{highlight1}{my-light-green}
\colorlet{highlight2}{my-yellow}
\colorlet{highlight3}{my-orange}
\colorlet{highlight4}{my-deep-orange!80!white}
\colorlet{highlight5}{my-yellow}
\colorlet{highlight6}{my-orange}
\colorlet{highlight7}{my-light-blue-grey}

\colorlet{highlightlight1}{highlight1!50!white}
\colorlet{highlightlight2}{highlight2!50!white}
\colorlet{highlightlight3}{highlight3!50!white}
\colorlet{highlightlight4}{highlight4!50!white}
\colorlet{highlightlight5}{highlight5!50!white}
\colorlet{highlightlight6}{highlight6!50!white}
\colorlet{highlightlight7}{highlight7!50!white}

\definecolor{foreground1}{HTML}{DD0000}
\definecolor{foreground2}{HTML}{00FF00}
\definecolor{foreground3}{HTML}{0000FF}
\usepackage{tikz} 
\usetikzlibrary{automata,arrows,arrows.meta,backgrounds,calc,decorations.markings,decorations.pathmorphing,decorations.pathreplacing,fit,positioning,shadows,shapes,shapes.geometric,shapes.misc,trees,patterns} 
\usetikzlibrary{pgfplots.groupplots}
\usepgfplotslibrary{fillbetween}

\newcommand{\psrc}[1]{\llap{\tikzmark{src#1}}}
\newcommand{\pdst}[1]{\llap{\tikzmark{dst#1}}}

\newcommand{\pdrawnext}[1]{{\tikz[overlay, remember picture] \draw[inner sep=0pt, outer sep=0pt] (src#1) ++(-.25em, -0.25em) edge[rounded corners=5mm, out=300, in=240, ->] ([xshift=-.25em, yshift=-0.25em]dst#1);%
}}

\newcommand{\pnode}[1]{{\tikz[overlay, remember picture] \node[draw=black, minimum width=1.4em, inner sep=2pt, xshift=-.25em] (pnode#1) {\scriptsize#1};}}

\newcommand{\pnodewl}[2]{\pnode{#1}{\tikz[overlay, remember picture] \node[below=-.4em of pnode#1] (pnodelabel#1) {\texttt{\footnotesize$\strut$#2}};}}

\newcommand{\pnodewls}[4]{%
    {\tikz[overlay, remember picture] \node[draw=black, minimum width=1.4em, inner sep=2pt, xshift=-.25em, #3] (pnode#1) {\scriptsize#1};}%
    {\tikz[overlay, remember picture] \node[below=-.4em of pnode#1, #4] (pnodelabel#1) {\texttt{\footnotesize$\strut$#2}};}}

\newcommand{\pnodewlhl}[3]{\pnodewls{#2}{#3}{fill=highlight#1, thick}{thick}}

\newcommand{\pedgeprev}[2]{{\tikz[overlay, remember picture] \draw[->, rounded corners=.5em] (pnode#1.north) |- (pnode#2.east);}}

\theoremseparator{.}
\theoremstyle{change}
\newcounter{allenvironments}
\newtheorem{theorem}{Theorem}[section]

\newtheorem{lemma}[theorem]{Lemma}
\newtheorem{corollary}[theorem]{Corollary}

\theoremstyle{nonumberplain}
\renewtheorem{theorem*}{Theorem}
\renewtheorem{satz*}{Satz}
\renewtheorem{lemma*}{Lemma}
\renewtheorem{corollary*}{Corollary}
\renewtheorem{proposition*}{Proposition}
\theorembodyfont{\upshape}
\theoremstyle{change}
\newmdtheoremenv[%
needspace=0\baselineskip,
outerlinewidth=1pt,
middlelinewidth=1pt,
innerlinewidth=1pt,
innerrightmargin=0pt,
innertopmargin=0pt,
innerbottommargin=0pt,
innerleftmargin=10pt,
leftmargin=-13pt,
skipabove=\baselineskip,
middlelinecolor=white,
bottomline=false,%
topline=false,%
rightline=false]{observationinternal}[theorem]{Observation}
\newmdtheoremenv[%
needspace=6\baselineskip,
outerlinewidth=1pt,
middlelinewidth=1pt,
innerlinewidth=1pt,
innerrightmargin=0pt,
innertopmargin=0pt,
innerbottommargin=0pt,
innerleftmargin=10pt,
leftmargin=-13pt,
skipabove=\baselineskip,
middlelinecolor=white,
bottomline=false,%
topline=false,%
rightline=false]{definitioninternal}[theorem]{Definition}
\newenvironment{definition}[1]{%
    \begin{center}\vspace{.5\baselineskip}\begin{definitioninternal}#1}{%
    \end{definitioninternal}\vspace{-.5\baselineskip}\end{center}%
}

\theoremstyle{nonumberplain}
\theoremheaderfont{\itshape}

\theoremsymbol{\ensuremath{\Box}}
\newtheorem{proof}{Proof}
\theoremsymbol{}
\theoremstyle{change}
\theoremheaderfont{\bfseries}

\theoremstyle{nonumberplain}
\renewtheorem{bemerkung*}{Bemerkung}
\renewtheorem{beispiel*}{Beispiel}
\renewtheorem{problem*}{Problem}

\usepackage[section]{algorithm}
\usepackage[noend]{algpseudocode}
\usepackage[displaymath, mathlines]{lineno}

\errorcontextlines\maxdimen

\newcommand{\ALGtikzmarkcolor}{black}
\newcommand{\ALGtikzmarkextraindent}{4pt}
\newcommand{\ALGtikzmarkverticaloffsetstart}{-.5ex}
\newcommand{\ALGtikzmarkverticaloffsetend}{-.5ex}
\makeatletter
\newcounter{ALG@tikzmark@tempcnta}

\newcommand\ALG@tikzmark@start{%
    \global\let\ALG@tikzmark@last\ALG@tikzmark@starttext%
    \expandafter\edef\csname ALG@tikzmark@\theALG@nested\endcsname{\theALG@tikzmark@tempcnta}%
    \tikzmark{ALG@tikzmark@start@\csname ALG@tikzmark@\theALG@nested\endcsname}%
    \addtocounter{ALG@tikzmark@tempcnta}{1}%
}

\def\ALG@tikzmark@starttext{start}
\newcommand\ALG@tikzmark@end{%
    \ifx\ALG@tikzmark@last\ALG@tikzmark@starttext
    \else
    \tikzmark{ALG@tikzmark@end@\csname ALG@tikzmark@\theALG@nested\endcsname}%
    \tikz[overlay,remember picture] \draw[\ALGtikzmarkcolor] let \p{S}=($(ALG@tikzmark@start@\csname ALG@tikzmark@\theALG@nested\endcsname)+(\ALGtikzmarkextraindent,\ALGtikzmarkverticaloffsetstart)$), \p{E}=($(ALG@tikzmark@end@\csname ALG@tikzmark@\theALG@nested\endcsname)+(\ALGtikzmarkextraindent,\ALGtikzmarkverticaloffsetend)$) in (\x{S},\y{S})--(\x{S},\y{E});%
    \fi
    \gdef\ALG@tikzmark@last{end}%
}

\apptocmd{\ALG@beginblock}{\ALG@tikzmark@start}{}{\errmessage{failed to patch}}
\pretocmd{\ALG@endblock}{\ALG@tikzmark@end}{}{\errmessage{failed to patch}}
\makeatother
\newcommand{\rungreek}{\mu}

\newcommand{\todoi}[1]{\todo[inline]{#1}}

\newcommand{\idx}[1]{\phantom{\texttt{0}}\llap{{\scriptsize#1}}}

\newcommand{\ellmax}{{\ell_\text{max}}}
\newcommand{\pmax}{{p_\text{max}}}

\newcommand{\unew}{{w}}

\newcommand{\flbox}[1]{\fbox[ltb]{#1\hspace*{-\fboxsep}}}
\newcommand{\frbox}[1]{\fbox[rtb]{\hspace*{-\fboxsep}#1}}
\newcommand{\fmbox}[1]{\fbox[tb]{\hspace*{-\fboxsep}#1\hspace*{-\fboxsep}}}

\newcommand{\strboxphant}{()}
\newcommand{\stridxphant}{\vphantom{\absolute{\mu}}}

\newlength{\defaultfboxsep}
\newcommand{\fboxcol}[2]{%
	\setlength{\fboxsep}{0pt}%
	\fbox{\setlength{\fboxsep}{\defaultfboxsep}\colorbox{#1}{#2}}%
	\setlength{\fboxsep}{\defaultfboxsep}%
}

\newcommand{\strboxhl}[3][\strboxphant]{\fboxcol{#2}{\ensuremath{\vphantom{#1}\smash{#3}}}}
\newcommand{\strlboxhl}[3][\strboxphant]{\flbox{\ensuremath{\vphantom{#1}\smash{#3}}}}
\newcommand{\strrboxhl}[3][\strboxphant]{\frbox{\ensuremath{\vphantom{#1}\smash{#3}}}}
\newcommand{\strmboxhl}[3][\strboxphant]{\fmbox{\ensuremath{\vphantom{#1}\smash{#3}}}}

\newcommand{\strbox}[2][\strboxphant]{\strboxhl[#1]{white}{#2}}
\newcommand{\strlbox}[2][\strboxphant]{\strlboxhl[#1]{white}{#2}}
\newcommand{\strrbox}[2][\strboxphant]{\strrboxhl[#1]{white}{#2}}
\newcommand{\strmbox}[2][\strboxphant]{\strmboxhl[#1]{white}{#2}}

\newcommand{\stridx}[2][]{\ensuremath{\overset{\mathclap{#2}\stridxphant}{\vphantom{\strbox{}}#1}}}
\newcommand{\stridxarrow}[2][]{\stridx[#1]{\overset{\scriptstyle#2\stridxphant}{\downarrow}}}

\newcommand{\rankopenplain}{\textnormal{rank}_{\textnormal{open}}}
\newcommand{\selectopenplain}{\textnormal{select}_{\textnormal{open}}}
\newcommand{\selectunclplain}{\textnormal{select}_{\textnormal{uncl}}}

\newcommand{\rankopen}[1]{\rankopenplain(#1)}
\newcommand{\selectopen}[1]{\selectopenplain(#1)}
\newcommand{\selectuncl}[1]{\selectunclplain(#1)}

\newcommand{\orderof}[1]{\mathcal{O}(#1)}
\newcommand{\orderofless}[1]{o(#1)}
\newcommand{\thetaof}[1]{\Theta(#1)}

\newcommand{\absolute}[1]{\left\lvert#1\right\rvert}
\newcommand{\floor}[1]{\left\lfloor#1\right\rfloor}
\newcommand{\ceil}[1]{\left\lceil#1\right\rceil}

\newcommand{\anchor}{\chi}
\newcommand{\xssfont}[1]{\textnormal{\textsf{#1}}}
\newcommand{\xssarg}[1]{[#1]}
\newcommand{\lyndarr}{\lambda}
\newcommand{\nss}[1]{{\xssfont{nss}%
\def\temp{#1}\ifx\temp\empty
\else
  \xssarg{#1}%
\fi}}
\newcommand{\pss}[1]{{\xssfont{pss}%
\def\temp{#1}\ifx\temp\empty
\else
  \xssarg{#1}%
\fi}}

\newcommand{\treeopfont}[1]{\textnormal{\textsc{#1}}}
\newcommand{\psstree}{\mathcal{T}_{\xssfont{pss}}}
\newcommand{\bpstree}{\mathcal{B}_{\xssfont{pss}}}
\newcommand{\pssbps}{\bpstree}
\newcommand{\parent}[1]{\treeopfont{Parent}(#1)}
\newcommand{\subtreesize}[1]{\treeopfont{SubtreeSize}(#1)}

\newcommand{\str}{\mathcal{S}}
\newcommand{\stralt}{\mathcal{T}}

\newcommand{\examplestr}[1]{\texttt{#1}}

\newcommand{\lce}[1]{\textnormal{\textsc{lce}}(#1)}
\newcommand{\lcp}[1]{\textnormal{\textsc{lcp}}(#1)}

\newcommand{\llex}{\prec}
\newcommand{\glex}{\succ}

\newcommand{\geqlex}{\succeq}

\newcommand{\algofont}[1]{\textsc{#1}}

\newcommand{\nssconcept}{\algofont{BuildPssBps}}

\newcommand{\pset}{\mathcal{P}}

\newcommand{\tikzol}[2][]{%
\begin{tikzpicture}[remember picture, overlay, #1]%
	#2%
\end{tikzpicture}%
}

\newcommand{\tikzmark}[2][]{%
	\tikzol{\node (tmp1) {};}%
	#1%
	\tikzol{
		\node (tmp2) {};
		\path (tmp1.center) edge[draw=none] node[midway, inner sep=0, outer sep=0] (#2) {} (tmp2.center);
	}%
}

\title{Space Efficient Construction of Lyndon Arrays\\in Linear Time}

\usepackage{authblk}

\author[1]{Philip Bille}
\author[2]{Jonas Ellert}
\author[2]{Johannes Fischer}
\author[1]{Inge Li G\o rtz}
\author[2]{Florian Kurpicz}
\author[3]{Ian Munro}
\author[1]{Eva Rotenberg}
\affil[1]{\small Department of Applied Mathematics and Computer Science, Technical University of Denmark}
\affil[2]{\small Department of Computer Science, Technical University of Dortmund, Germany}
\affil[3]{\small Cheriton School of Computer Science, University of Waterloo, Canada}

\date{\today}

\begin{document}

\maketitle

\begin{abstract}
We present the first linear time algorithm to construct the $2n$-bit version of the Lyndon array for a string of length $n$ using only $\orderofless{n}$ bits of working space.
A simpler variant of this algorithm computes the plain ($n\lg n$-bit) version\footnote{Throughout this article, $\lg$ denotes the binary logarithm} of the Lyndon array using only $\orderof{1}$ words of additional working space.
All previous algorithms are either not linear, or use at least $n\lg n$ bits of additional working space.
Also in practice, our new algorithms outperform the previous best ones by an order of magnitude, both in terms of time and space.
\end{abstract}

\section{Introduction \& Related Work}

The suffix array is one of the most important data structures in the field of string processing. Baier's suffix array construction algorithm was the first to compute the suffix array in linear time without using recursion \citep{Baier2016}. Interestingly, in its first phase the algorithm computes a partially sorted version of the Lyndon array \citep{Franek2017}. While it was previously known that the Lyndon array can be computed in linear time from the suffix array \citep{Hohlweg2003,Franek2016}, Baier's algorithm is the only one that computes the Lyndon array as a preliminary data structure for computing the suffix array. However, even for the first phase --- and thus for computing the Lyndon array --- it requires $\Theta(n \lg n)$ bits of additional working space. Since then, multiple algorithms for more space efficient Lyndon array construction have been introduced (e.g.\ \citep{Franek2016,Louza2018,Louza2019}). However, all linear time approaches either have to precompute the suffix array, or they simultaneously compute the Lyndon array and the suffix array, which means that they need at least $n \ceil{\lg n}$ bits of working space to store the suffix array. For example, the currently best known result uses ${\sigma} \ceil{\lg n}$ bits of additional working space to compute the Lyndon array and the suffix array in linear time, and thus $(n + \sigma) \ceil{\lg n}$ bits of additional working space to compute the Lyndon array \citep{Louza2019}. 

The Lyndon array has some structural properties that allow for a more space efficient representation, namely using only $2n+2$ bits \citep{Louza2018}. Thus, it would be desirable to compute this succinct representation using less than $\Theta(n\lg n)$ bits of working space, without sacrificing the linear running time. Currently, no such algorithm exists.

\subsection{Our Contributions}
We introduce the first algorithm that computes the succinct Lyndon array 
in $\orderof{n}$ time, using only $\orderof{n / \lg\lg n}$ bits of additional working space. Alternatively, our algorithm can also construct the plain ($\orderof{n \lg n}$-bits) Lyndon array using only $\orderof{1}$ words of additional working space, i.e., directly without having to precompute the the suffix array.

 The algorithm is not only the currently best solution in terms of theoretical worst case bounds, but also almost 10 times faster than other state of the art Lyndon array construction algorithms in practice.

The paper is structured as follows: First, we propose a more intuitive definition of the succinct Lyndon array representation from \citep{Louza2018}. Then, we introduce our construction algorithm, which directly computes the succinct representation. Finally, we adapt the algorithm such that it computes the Lyndon array in its naive representation and present experimental results for both versions.

\section{Preliminaries}

Since we only use logarithms to base two, we simply write $\lg x$ instead of $\log_2 x$. All intervals are to be considered discrete, i.e.\ for $i,j \in \mathbb{N}$ the interval $[i,j]$ represents the set ${\{x \mid x \in \mathbb{N} \land i \leq x \leq j \}}$. We use the notation $[i, j + 1) = (i - 1, j] = (i-1, j+1) = [i, j]$ for open and half-open discrete intervals. Our research is situated in the word RAM model \citep{Hagerup1998}, where we can perform fundamental operations (logical shifts, basic arithmetic operations etc.) on words of size $w$ bits in constant time. For the input size $n$ of our problems we assume $\ceil{\lg n} \leq w$.

A \emph{string} (also called \emph{text}) over the \emph{alphabet} $\Sigma$ is a finite sequence of \emph{symbols} from the finite and totally ordered set $\Sigma$. We say that a string $\str$ has length $n$ and write $\absolute{\str} = n$, iff $\str$ is a sequence of exactly $n$ symbols. 
The $i$-th symbol of a string $\str$ is denoted by $\str[i]$, while the \emph{substring} from the $i$-th to the $j$-th symbol is denoted by $\str[i..j]$. 
For convenience we use the interval notations $\str[i..j + 1) = \str(i - 1..j] = \str(i-1..j+1) = \str[i..j]$. The $i$-th \emph{suffix} of $\str$ is defined as $\str_i = \str[i..n]$, while the substring $\str[1..i]$ is called \emph{prefix} of $\str$. A prefix or suffix of $\str$ is called \emph{proper}, iff its length is at most $n - 1$. The concatenation of two strings $\str$ and $\stralt$ is denoted by $\str \cdot \stralt$. The length of the \emph{longest common prefix (LCP)} between $\str$ and $\stralt$ is defined as $\lcp{\str, \stralt} = \max\{ \ell \mid \str[1..\ell] = \stralt[1..\ell]\}$. The \emph{longest common extension (LCE)} of indices $i$ and $j$ is the length of the LCP between $\str_i$ and $\str_j$, i.e.\ $\lce{i, j} = \lcp{\str_i, \str_j}$.
	
We can simplify the description of our algorithm by introducing a special symbol $\examplestr{\$} \notin \Sigma$ that is smaller than all symbols from $\Sigma$. For a string $\str$ of length $n$ we define the \emph{$0$-th suffix} $\str_0 = \examplestr{\$}$ as well as the \emph{$(n + 1)$-th suffix and position} $\str_{n + 1}= \str[n + 1] = \examplestr{\$}$. The total order on $\Sigma$ induces a total order on the set $\Sigma^*$ of strings over $\Sigma$. Let $\str$ and $\stralt$ be strings over $\Sigma$, and let $\ell = \lcp{\str, \stralt}$. We say that $\str$ is lexicographically smaller than $\stralt$ and write $\str \llex \stralt$, iff we have $\str \neq \stralt$ and $\str[\ell + 1] < \stralt[\ell + 1]$. Analogously, we say that $\str$ is lexicographically larger than $\stralt$ and write $\str \glex \stralt$, iff we have $\str \neq \stralt$ and $\str[\ell + 1] > \stralt[\ell + 1]$.


\subsection{The Lyndon Array \& Nearest Smaller Suffixes}

A \emph{Lyndon word} is a string that is lexicographically smaller than all of its proper prefixes, i.e.\ $\str$ is a Lyndon word, iff $\forall i \in [2..n] : \str_i \llex \str$ holds \citep{Duval1983}. For example, the string $\examplestr{northamerica}$ is not a Lyndon word because its suffix $\examplestr{america}$ is lexicographically smaller than itself. On the other hand, its substring $\examplestr{americ}$ is a Lyndon word. The Lyndon array of $\str$ identifies the longest Lyndon word that begins at each position of $\str$:

\begin{definition}[Lyndon Array]
	Given a string $\str$ of length $n$, the Lyndon array is an array $\lyndarr$ of $n$ integers with $\lyndarr[i] = \max\{ \ell \mid \str[i..i+\ell) \text{ is a Lyndon word} \}$.
\end{definition}
\begin{definition}[Nearest Smaller Suffixes]\label{def:xss}
	Given a string $\str$ and a suffix $\str_i$, the \emph{next smaller suffix of $\str_i$} is $\str_j$, where $j$ is the smallest index that is larger than $i$ and satisfies $\str_i \glex \str_j$. The \emph{previous smaller suffix of $\str_i$} is defined analogously. The \emph{next and previous smaller suffix arrays} are arrays of size $n$ that are defined as follows:
	\begin{center}
	$\nss{i} = \min\{j \mid j \in (i, n + 1] \land \str_i \glex \str_j \} \qquad \pss{i} = \max \{j \mid j \in [0, i) \land \str_j \llex \str_i\}$
	\end{center}
\end{definition}

\noindent The Lyndon array and nearest smaller suffixes are highly related to each other. In fact, the next smaller suffix array is merely a different representation of the Lyndon array: 

\begin{lemma}[Lemma 15 \citep{Franek2016}]\label{lemma:lyndonnss}
	The longest Lyndon word at position $i$ ends at the starting position of the NSS of $\str_i$, i.e.\ ${\lyndarr[i] = \nss{i} - i}$.
\end{lemma}

A visualization of the Lyndon array and the NSS array can be found in \cref{fig:lyndonexample}. We conclude the preliminaries by showing a slightly weaker connection between the PSS array and Lyndon words:

\Needspace{5\baselineskip}\begin{lemma}
	Let $\pss{j} = i > 0$, then $\str[i..j)$ is a Lyndon word.
	\begin{proof}
		By definition, the string $\str[i..j)$ is a Lyndon word iff there exists no ${k \in (i, j)}$ with ${\str[k..j) \llex \str[i..j)}$. Assume that such a $k$ exists.
        Since $i = \pss{j}$, we know that (a) $\str_k \glex \str_i$.
        Now assume there is a mismatching character between $\str[k..j)$ and $\str[i..j)$.
        Then appending $\str_j$ to both strings preserves this mismatch.
        This implies that we have ${\str[k..j) < \str[i..j)} \Longleftrightarrow {\str[k..j)\cdot \str_j < \str[i..j)\cdot \str_j}$, and thus $\str_k \llex \str_i$, which contradicts (a).
        Therefore, we know that (b) $\str[k..j) = \str[i..i + (j - k))$.
        Then 
        \begin{alignat*}{4}
        &\ \ &\str_k\qquad\overset{\text{(a)}}{\glex}&\qquad\str_i\\
        \Longleftrightarrow &&\str[k..j)\cdot \str_j\enskip\glex&\enskip\str[i..i + (j - k))\cdot \str_{i + (j - k)}\\
        \underset{\text{(b)}}{\Longleftrightarrow} &&\str_j\qquad\glex&\qquad\str_{i + (j - k)}\ ,
        \end{alignat*}
        which contradicts the fact that $\pss{j} = i < i+(j-k)$.
        Hence, the described $k$ cannot exist, and $\str[i..j)$ must be a Lyndon word.
	\end{proof}
	\label{lemma:psslyndonword}
\end{lemma}

\section{Previous Smaller Suffix Trees}
\label{sec:psstrees}
\begin{figure}[t]
    \centering
    \small
	\begin{subfigure}[t]{.42\textwidth}
	    \newcommand{\namericaphantom}{$\strut$\phantom{$\nss{} \ =\ $\texttt{\ n o r t h a m e r i c a \$}}}
	    \centering
        \namericaphantom\llap{$\strut%
            \idx1%
            \texttt{ }\idx2%
            \texttt{ }\idx3%
            \texttt{ }\idx4%
            \texttt{ }\idx5%
            \texttt{ }\idx6%
            \texttt{ }\idx7%
            \texttt{ }\idx8%
            \texttt{ }\idx9%
            \texttt{ }\idx{10}%
            \texttt{ }\idx{11}%
            \texttt{ }\idx{12}%
            \texttt{ \ }$%
        }\\[-.25em]
        \namericaphantom\llap{$\str \ =\ $\texttt{\ n o r t h a m e r i c a \ }}\\[-.25em]
        \namericaphantom\llap{$\strut\lyndarr \ =\ $%
            $\texttt{ }\idx4%
            \texttt{ }\idx3%
            \texttt{ }\idx2%
            \texttt{ }\idx1%
            \texttt{ }\idx1%
            \texttt{ }\idx6%
            \texttt{ }\idx1%
            \texttt{ }\idx3%
            \texttt{ }\idx1%
            \texttt{ }\idx1%
            \texttt{ }\idx1%
            \texttt{ }\idx1%
            \texttt{ \ }$%
        }\\[-.25em]
        \namericaphantom\llap{$\strut\nss{} \ =\ $%
            $\texttt{ }\idx5\psrc{1}%
            \texttt{ }\idx5\psrc{2}%
            \texttt{ }\idx5\psrc{3}%
            \texttt{ }\idx5\psrc{4}%
            \texttt{ }\idx6\pdst{1}\pdst{2}\pdst{3}\pdst{4}\psrc{5}%
            \texttt{ }\idx{12}\pdst{5}\psrc{6}%
            \texttt{ }\idx8\psrc{7}%
            \texttt{ }\idx{11}\pdst{7}\psrc{8}%
            \texttt{ }\idx{10}\psrc{9}%
            \texttt{ }\idx{11}\pdst{9}\psrc{10}%
            \texttt{ }\idx{12}\pdst{10}\pdst{8}\psrc{11}
            \texttt{ }\idx{13}\pdst{11}\pdst{6}\psrc{12}%
            \texttt{ \ }\pdst{12}$%
        }\\        
        \pdrawnext{1}%
        \pdrawnext{2}%
        \pdrawnext{3}%
        \pdrawnext{4}%
        \pdrawnext{5}%
        \pdrawnext{6}%
        \pdrawnext{7}%
        \pdrawnext{8}%
        \pdrawnext{9}%
        \pdrawnext{10}%
        \pdrawnext{11}%
        \pdrawnext{12}%
        \\[.25\baselineskip]\newcommand{\lyndonspace}{-.4\baselineskip}
        \namericaphantom\llap{\texttt{n o r t \ \ \ \ \ \ \ \ \ \ \ \ \ \ \ \ \ }}\\[\lyndonspace]
        \namericaphantom\llap{\texttt{  o r t \ \ \ \ \ \ \ \ \ \ \ \ \ \ \ \ \ }}\\[\lyndonspace]
        \namericaphantom\llap{\texttt{    r t \ \ \ \ \ \ \ \ \ \ \ \ \ \ \ \ \ }}\\[\lyndonspace]
        \namericaphantom\llap{\texttt{      t \ \ \ \ \ \ \ \ \ \ \ \ \ \ \ \ \ }}\\[\lyndonspace]
        \namericaphantom\llap{\texttt{        h \ \ \ \ \ \ \ \ \ \ \ \ \ \ \ }}\\[\lyndonspace]
        \namericaphantom\llap{\texttt{         a m e r i c \ \ \ }}\\[\lyndonspace]
        \namericaphantom\llap{\texttt{           m \ \ \ \ \ \ \ \ \ \ \ }}\\[\lyndonspace]
        \namericaphantom\llap{\texttt{             e r i \ \ \ \ \ }}\\[\lyndonspace]
        \namericaphantom\llap{\texttt{               r \ \ \ \ \ \ \ }}\\[\lyndonspace]
        \namericaphantom\llap{\texttt{                 i \ \ \ \ \ }}\\[\lyndonspace]
        \namericaphantom\llap{\texttt{                   c \ \ \ }}\\[\lyndonspace]
        \namericaphantom\llap{\texttt{                     a \ }}
        \caption{Lyndon array, NSS array, and maximal Lyndon words at all indices of $\str$.}
        \label{fig:lyndonexample}
	\end{subfigure}
	\hspace{.03\textwidth}
    \begin{subfigure}[t]{.52\textwidth}
    \renewcommand{\psrc}[1]{\llap{\tikzmark{src#1}}}
	\renewcommand{\pdst}[1]{\llap{\tikzmark{dst#1}}}
	\newcommand{\llaptt}[1]{{\phantom{\texttt{0}}}\llap{#1}}
	\newcommand{\rlaptt}[1]{\rlap{#1}{\phantom{\texttt{0}}}}
	\renewcommand{\idx}[1]{\llaptt{{\scriptsize#1}}}
	\newcommand{\idxhl}[2]{\llaptt{\colorbox{highlight#1}{\scriptsize#2}\kern-.25em}}    
	\newcommand{\opclallkern}{\kern.4em}
\newcommand{\opkern}{\opclallkern}
\newcommand{\clkern}{\opclallkern}
\newcommand{\opclkern}{\opclallkern}

\newcommand{\opbg}[1]{\rlap{\kern-.275em{\colorbox{#1}{\texttt{\phantom{)}}}}}\texttt{\textbf{(}}}
\newcommand{\clbg}[1]{\rlap{\kern-.275em{\colorbox{#1}{\texttt{)}}}}\texttt{ }}
\newcommand{\opclbg}[2]{\rlap{\kern-.275em{\colorbox{#1}{\texttt{\textbf{(}}\tikzmark{dstop#2}\texttt{)}}}}\texttt{ \ }}

\newcommand{\op}[1]{\tikzmark{srcop#1}\ensuremath{\overset{\mathclap{{\scriptsize#1}}}{\texttt{\textbf{(}}}}\opkern}
\newcommand{\cl}[1]{\tikzmark{dstop#1}\ensuremath{\underset{\mathclap{{\phantom{\scriptsize#1}}}}{\texttt{)}}}\clkern}
\newcommand{\opcl}[1]{\tikzmark{srcop#1}\ensuremath{\overset{\mathclap{{\scriptsize#1}}}{\texttt{\textbf{(}}\tikzmark{dstop#1}\texttt{)}}}\opclkern}

\newcommand{\ophl}[2]{\tikzmark{srcop#2}\ensuremath{\overset{\mathclap{{\scriptsize#2}}}{\opbg{highlight#1}}}\opkern}
\newcommand{\clhl}[2]{\tikzmark{dstop#2}\ensuremath{\underset{\mathclap{{\phantom{\scriptsize#2}}}}{\clbg{highlight#1}}}\clkern}
\newcommand{\opclhl}[2]{\tikzmark{srcop#2}\ensuremath{\overset{\mathclap{{\scriptsize#2}}}{\opclbg{highlight#1}{#2}}}\opclkern}

\newcommand{\opclunderlinescale}{.4}
\newcommand{\opclunderline}[2]{%
    \tikz[overlay, remember picture] \draw[transform canvas={xshift=0.4em, yshift={(-#1*\opclunderlinescale - .3)*1em}}, thick, -{|[width=.35em]}] (srcop#2) ++(-.1em, {(#1*\opclunderlinescale)*1em}) |- (dstop#2);%
}

\newcommand{\opclunderlinehl}[3]{%
    \tikz[overlay, remember picture] \fill[highlight#1, transform canvas={xshift=0.35em, yshift={(-#2*\opclunderlinescale - .3)*1em}}, -{|[width=.35em]}] (srcop#3) ++(-.05em, {(#2*\opclunderlinescale)*1em}) rectangle (dstop#3);%
    \tikz[overlay, remember picture] \draw[thick, transform canvas={xshift=0.4em, yshift={(-#2*\opclunderlinescale - .3)*1em}}, -{|[width=.35em]}] (srcop#3) ++(-.1em, {(#2*\opclunderlinescale)*1em}) |- (dstop#3);%
}
    \newcommand{\namericaphantom}{$\strut$\phantom{$\pss{} \ =\ $\texttt{\ n \ o \ r \ t \ h \ a \ m \ e \ r \ i \ c \ a \ \ }}}
    \namericaphantom\llap{$\strut%
        \idx1%
        \texttt{ \ }\idx2%
        \texttt{ \ }\idx3%
        \texttt{ \ }\idx4%
        \texttt{ \ }\idx5%
        \texttt{ \ }\idx6%
        \texttt{ \ }\idx7%
        \texttt{ \ }\idx8%
        \texttt{ \ }\idx9%
        \texttt{ \ }\idx{10}%
        \texttt{ \ }\idx{11}%
        \texttt{ \ }\idx{12}%
        \texttt{ \ \ }$%
    }\\[-.25em]
    \namericaphantom\llap{$\str \ =\ $\texttt{\ n \ o \ r \ t \ h \ a \ m \ e \ r \ i \ c \ a \ \ }}\\[-.25em]
    \namericaphantom\llap{$\strut\pss{} \ =\ $%
        $\texttt{\ }\idx0\pdst{13}\pdst{12}\pdst{6}\pdst{5}\pdst{1}%
        \texttt{ \ }\idx1\pdst{3}\psrc{1}%
        \texttt{ \ }\idx2\pdst{4}\psrc{2}%
        \texttt{ \ }\idx3\pdst{5}\psrc{3}%
        \texttt{ \ }\idx0\psrc{4}%
        \texttt{ \ }\idx0\psrc{5}%
        \texttt{ \ }\idxhl16\pdst{12}\pdst{9}\pdst{8}\psrc{6}%
        \texttt{ \ }\idxhl16\psrc{7}%
        \texttt{ \ }\idxhl28\pdst{11}\pdst{10}\psrc{8}%
        \texttt{ \ }\idxhl28\psrc{9}%
        \texttt{ \ }\idxhl16\psrc{10}%
        \texttt{ \ }\idx0\psrc{11}%
        \texttt{ \ \ }$%
    }\\[.25em]
    \namericaphantom\llap{%
    \phantom{\texttt{........................................}}}\\[-1.5em]
    \namericaphantom\llap{%
        \pnodewl{0}{\$}%
        \phantom{\texttt{..2..3..4..5..6..7..8..9..0..1..2..3..4}}%
    }\\
    \namericaphantom\llap{%
        \pnodewl{1}{n}%
        \phantom{\texttt{..3..4..5..6}}%
        \pnodewl{5}{h}%
        \phantom{\texttt{..7}}%
        \pnodewlhl1{6}{a}%
        \phantom{\texttt{..8..9..0..1..2..3}}%
        \pnodewl{12}{a}%
        \phantom{\texttt{..4}}%
    }\\
    \namericaphantom\llap{%
        \pnodewl{2}{o}%
        \phantom{\texttt{..4..5..6..7..8}}%
        \pnodewlhl2{7}{m}%
        \phantom{\texttt{..9}}%
        \pnodewlhl2{8}{e}%
        \phantom{\texttt{..0..1..2}}%
        \pnodewlhl2{11}{c}%
        \phantom{\texttt{..3..4}}%
    }\\
    \namericaphantom\llap{%
        \pnodewl{3}{r}%
        \phantom{\texttt{..5..6..7..8..9..0}}%
        \pnodewlhl3{9}{r}%
        \phantom{\texttt{..1}}%
        \pnodewlhl3{10}{i}%
        \phantom{\texttt{..2..3..4}}%
    }\\
    \namericaphantom\llap{%
        \pnodewl{4}{t}%
        \phantom{\texttt{..6..7..8..9..0..1..2..3..4}}%
    }\\
    \pedgeprev{1}{0}%
    \pedgeprev{5}{0}%
    \pedgeprev{6}{0}%
    \pedgeprev{12}{0}%
    \pedgeprev{2}{1}%
    \pedgeprev{7}{6}%
    \pedgeprev{8}{6}%
    \pedgeprev{11}{6}%
    \pedgeprev{3}{2}%
    \pedgeprev{9}{8}%
    \pedgeprev{10}{8}%
    \pedgeprev{4}{3}\\[.25\baselineskip]
    \namericaphantom\llap{%
    \phantom{\texttt{........................................}}}\llap{%
        $\op{0}\op{1}\op{2}\op{3}\opcl{4}\cl{3}\cl{2}\cl{1}\opcl{5}\ophl1{6}\opclhl2{7}\ophl2{8}\opclhl3{9}\opclhl3{10}\clhl2{8} \opclhl2{11}\clhl1{6}\opcl{12}\cl{0}\ $}\\[1.2\baselineskip]
    \opclunderlinehl1{4}{6}%
    \opclunderlinehl2{3}{7}%
    \opclunderlinehl2{3}{8}%
    \opclunderlinehl3{2}{9}%
    \opclunderlinehl3{2}{10}%
    \opclunderlinehl2{3}{11}%
    \caption{PSS array, PSS tree, and BPS representation of the PSS tree of $\str$ (best viewed in color).}
    \label{fig:psstreebps}
    \end{subfigure}
    \caption{Data structures for $\str = \examplestr{northamerica}$.}
\end{figure}

In this section we introduce the \emph{previous smaller suffix tree}, which simulates access to the Lyndon array, the NSS array, and the PSS array. The PSS array inherently forms a tree in which each index $i$ is represented by a node whose parent is $\pss{i}$. The root is the artificial index $0$, which is parent of all indices that do not have a previous smaller suffix (see \cref{fig:psstreebps} for an example). 
\begin{definition}[Previous Smaller Suffix Tree $\psstree$]
    Let $\str$ be a string of length $n$. The \emph{previous smaller suffix tree (PSS tree)} of $\str$ is an ordinal tree $\psstree$ with nodes $[0, n]$ and root $0$. For $i \in [1, n]$, we define $\parent{i} = \pss{i}$. The children are ordered ascendingly, i.e.\ if $i$ is a left-side sibling of $j$, then $i < j$ holds.
    \label{def:psstree}
\end{definition}%
The PSS tree is highly similar to the LRM tree \citep{Sadakane2010, Fischer2010}. In fact, the only difference is that in the LRM tree each index is attached to its previous smaller \emph{value} instead of its previous smaller \emph{suffix}. Since the NSS array of a string is identical to the NSV array of its inverse suffix array (see e.g.\ \citep[Algorithm NSVISA]{Franek2016}), it follows that the PSS tree of a string is identical to the LRM tree of its inverse suffix array. Consequently, an important property of the LRM tree also applies to the PSS tree:
\begin{corollary}[Lemma 1 \citep{Fischer2010}]
	The nodes of the PSS tree directly correspond to the preorder-numbers, i.e.\ node $i$ has preorder-number $i$ (assuming that the first preorder-number is 0).
	\label{lemma:treepreorder}
\end{corollary}
The corollary allows us to simulate the NSS array with the PSS tree:
\begin{lemma}
	Given the PSS tree, NSS array and Lyndon array of the same string we have $\nss{i} = i + \subtreesize{i}$ and thus $\lyndarr[i] = \subtreesize{i}$.
	\begin{proof}
		Since the nodes directly correspond to the preorder-numbers (\cref{lemma:treepreorder}), it follows that the descendants of $i$ form a consecutive interval $(i, r]$. Since $i + \subtreesize{i} = i + (r - i + 1) = r + 1$ holds, we only have to show $\nss{i} = r + 1$. Assume $r = n$, then there is no index larger than $i$ that is not a descendant of $i$. Clearly, in this case $i$ does not have an NSS, and thus it follows $\nss{i} = n + 1 = r + 1$. Assume $r < n$ instead, then $\str_{r + 1}$ must be lexicographically smaller than all suffixes that begin at positions from $[i, r]$, since otherwise $r + 1$ would be a descendant of $i$. Therefore, $\str_{r + 1}$ is the first suffix that starts right of $i$ and is lexicographically smaller than $\str_i$, which means $\nss{i} = r + 1$.
	\end{proof}
	\label{lemma:nssfrompsstree}
\end{lemma}
Note in particular that since different suffixes of a text cannot be equal, the data structure needs only $2n$ bits instead of the $2.54\dots n$ as in the case of previous and next smaller \textit{values} \citep{Fischer2011}.

\subsection{Storing the PSS Tree as a BPS}

We store the PSS tree as a balanced parentheses sequence (BPS, \citep{Munro2001}) of length $2n + 2$, which takes $2n + 2$ bits. As a shorthand for the BPS of the PSS tree we write $\pssbps$. The sequence is algorithmically defined by a preorder-traversal of the PSS tree, where we write an opening parenthesis whenever we \emph{walk down} an edge, and a closing one whenever we \emph{walk up} an edge. An example is provided in \cref{fig:psstreebps}. Note that the BPS of the PSS tree is identical to the succinct Lyndon array presentation from \citep{Louza2018}.

While the BPS itself does not support fast tree operations, we can use it in combination with the data structure from \citep{Sadakane2010}, which takes $\orderof{n}$ time and $\orderof{n}$ bits of working space. This data structure is of size $\orderof{n / \lg^c n}$ bits (for any $c \in \mathbb{N}^{+}$ of our choice) and supports $\parent{\cdot}$ and $\subtreesize{\cdot}$ operations in $\orderof{c^2}$ time, and allows us to simulate access to the Lyndon array in $\orderof{c^2}$ time using \cref{lemma:nssfrompsstree}.

\paragraph{Operations on a BPS Prefix.}\label{sec:auxmaintain} Since we will be building $\pssbps$ from left to right, at any given point of the algorithm execution we know a prefix of $\pssbps$. It is crucial that we maintain support for the following queries in constant time:

\begin{itemize}	
	\item Given the index $o_i$ of an opening parenthesis in $\pssbps$, determine the node $i$ that belongs to the parenthesis. We have $i = \rankopen{o_i} - 1$, where $\rankopen{o_i}$ is the number of opening parentheses in $\pssbps[1..o_i]$.
	
	\item Given a preorder-number $i$, find the index $o_i$ of the corresponding opening parenthesis in $\pssbps$. We have $o_i = \selectopen{i} = \min\{o \mid \rankopen{o} > i\}$.
	
	\item Given an integer $k \geq 1$, find the index $o_{\textnormal{uncl}(k)} = \selectuncl{k}$ of the $k$-th \emph{unclosed} parenthesis in $\pssbps$. An opening parenthesis is called unclosed, if we have not written the matching closing parenthesis yet. For example, there are 5 opening parentheses in $(()(()$, but only the first and the third one are unclosed.
\end{itemize}

\noindent We can maintain support data structures of $\orderof{n \lg \lg n/ \lg n}$ bits to answer these queries in constant time \citep{Golynski2007}. The structure for $\selectunclplain$ is a trivial modification of the structure for $\selectopenplain$. Since this data structure can be constructed in linear time by scanning the BPS from left to right, clearly we can maintain them with no significant time overhead when writing the BPS in an append-only manner.

\section{Constructing the PSS Tree}

In this section we introduce our construction algorithm for the BPS of the PSS tree, which processes the indices of the input text in left-to-right order. Processing index $i$ essentially means that we attach $i$ to a partial PSS tree that is induced by the nodes from $[0, i)$. An example is provided in \cref{fig:exampleproci} (left). But how can we efficiently determine $pss[i]$, which is $i$'s parent? Consider the nodes on the rightmost path of the partial tree, which starts at $i - 1$ and ends at the root $0$. We call the set of these nodes \emph{PSS closure $\pset_{i - 1}$} of $i - 1$ because it contains exactly the nodes that can be obtained by repeated application of the PSS function on $i - 1$. For $j \in [1, n]$ we recursively define $\pset_0 = \{0\}$ and $\pset_{j} = \{j\} \cup \pset_{\pss{j}}$. Interestingly, $\pss{i}$ is a member of $\pset_{i - 1}$:

\begin{figure}
\centering
\small
\tikzset{pedgestyle/.style={->, rounded corners=.5em}}
\tikzset{pmaybestyle/.style={pedgestyle, dashed}}
\tikzset{hlnode/.style={fill=highlight1}}
\newcommand{\parentedge}[3][]{\draw[pedgestyle, #1] (#2.north) |- (#3.east);}
\newcommand{\maybeedge}[3][]{\draw[pmaybestyle, #1] (#2.north) |- (#3.east);}
\newcommand{\shortedge}[3][]{\draw[pedgestyle, #1] (#2.north east) |- (#3.east);}
\newcommand{\opclallkern}{\kern.3em}
\newcommand{\opkern}{\opclallkern}
\newcommand{\clkern}{\opclallkern}
\newcommand{\opclkern}{\opclallkern}
\newcommand{\opbg}[1]{\rlap{\kern-.275em{\colorbox{#1}{\texttt{\phantom{)}}}}}\texttt{\textbf{(}}}
\newcommand{\clbg}[1]{\rlap{\kern-.275em{\colorbox{#1}{\texttt{)}}}}\texttt{ }}
\newcommand{\op}[1]{\ensuremath{\overset{\mathclap{\phantom{\scriptsize#1}}}{\texttt{\textbf{(}}}}\opkern}
\newcommand{\cl}[1]{\ensuremath{\underset{\mathclap{{\phantom{\scriptsize#1}}}}{\texttt{)}}}\clkern}
\newcommand{\opcl}[1]{\ensuremath{\overset{\mathclap{\phantom{\scriptsize#1}}}{\texttt{\textbf{(}}\texttt{)}}}\opclkern}
\newcommand{\ophl}[2]{\ensuremath{\overset{\mathclap{{\scriptsize#2}}}{\opbg{highlight#1}}}\opkern}
\newcommand{\clhl}[2]{\ensuremath{\underset{\mathclap{{{\scriptsize#2}}}}{\clbg{highlight#1}}}\clkern}
\newcommand{\opclhl}[2]{\ensuremath{\overset{\mathclap{{\scriptsize#2}}}{\opclbg{highlight#1}{#2}}}\opclkern}
\begin{tikzpicture}[every node/.style={draw=black}]
\draw node[hlnode] (0) {$0$}
++(1.5, -1) node[hlnode] (6) {$6$}
++(.75, -1) node (7) {$7$} ++(.75, 0) node[hlnode] (8) {$8$}
++(.75, -1) node (9) {$9$} ++(.75, 0) node[hlnode] (10) {$10$}
++(1.5, 0) node (11) {$11$};
\draw (0) 
++(.25, -.6) node (1) {$1$}
++(.25, -.6) node (2) {$2$}
++(.25, -.6) node (3) {$3$}
++(.25, -.6) node (4) {$4$}
++(.25, -.6) node (5) {$5$};
\parentedge{6}{0};
\parentedge{7}{6};
\parentedge{8}{6};
\parentedge{9}{8};
\parentedge{10}{8};
\draw (11.west) edge[dashed, ->] node[draw=none, above] {?} (10.east);
\draw[dashed] (11.north) edge[out=90, in=0, ->] node[draw=none, below] {?} (10.center |- 8);
\draw[dashed] (11.north) edge[out=90, in=0, ->] node[draw=none, below] {?} (8.center |- 6);
\draw[dashed] (11.north) edge[out=90, in=0, ->] node[draw=none, below] {?} (6.center |- 0);
\shortedge{5}{4};
\shortedge{4}{3};
\shortedge{3}{2};
\shortedge{2}{1};
\shortedge{1}{0};
\node[draw = none] (bps) at ($(0 |- 5) + (0, -1)$) {\rlap{%
$\pssbps = \ophl1{0}\op{1}\op{2}\op{3}\opcl{4}\cl{3}\cl{2}\cl{1}\opcl{5}\ophl1{6}\opcl{7}\ophl1{8}\opcl{9}\ophl1{10}\phantom{\clhl2{10}}$%
}};
\end{tikzpicture}~~~~~~~~~
\begin{tikzpicture}[every node/.style={draw=black}]
\draw node (0) {$0$}
++(1.5, -1) node (6) {$6$}
++(.75, -1) node (7) {$7$} ++(.75, 0) node[fill=highlight2] (8) {$8$}
++(.75, -1) node (9) {$9$} ++(.75, 0) node[fill=highlight2] (10) {$10$}
++(1.5, 0) node[fill=highlight3] (11) {$11$} ++(1, 0);
\draw (0) 
++(.25, -.6) node (1) {$1$}
++(.25, -.6) node (2) {$2$}
++(.25, -.6) node (3) {$3$}
++(.25, -.6) node (4) {$4$}
++(.25, -.6) node (5) {$5$};
\parentedge{6}{0};
\parentedge{7}{6};
\parentedge[thick]{8}{6};
\parentedge{9}{8};
\parentedge[thick]{10}{8};
\parentedge[thick]{11}{6};
\shortedge{5}{4};
\shortedge{4}{3};
\shortedge{3}{2};
\shortedge{2}{1};
\shortedge{1}{0};
\node[draw = none] (bps) at ($(0 |- 5) + (0, -1)$) {\rlap{%
$\pssbps = \ophl1{0}\op{1}\op{2}\op{3}\opcl{4}\cl{3}\cl{2}\cl{1}\opcl{5}\ophl1{6}\opcl{7}\ophl1{8}\opcl{9}\ophl1{10}\clhl2{10}\clhl2{8}\ophl3{11}$%
}};
\end{tikzpicture}
\caption{The partial PSS tree before (left) and after (right) processing index $11$ of $\str = \examplestr{northamerica}$ during the execution of \cref{alg:concept}. We have $p_1 = 10$, $p_2 = 8$, $p_3 = 6$, $p_4 = 0$, and $p_m = p_3$.}
\label{fig:exampleproci}
\end{figure}

\begin{lemma}
For any index $i \in [1, n]$ we have $\pss{i} = \max \{j \mid j \in \pset_{i - 1} \land \str_j \llex \str_i\}$.
\label{lemma:pssinclosure}
\begin{proof}
	If we show $\pss{i} \in \pset_{i - 1}$, then the correctness of the lemma follows from \cref{def:xss}. Assume $\pss{i} \notin \pset_{i - 1}$, then there is some index $j \in \pset_{i - 1}$ with $\pss{i} \in (\pss{j}, j)$. By \cref{def:xss}, this implies $\str_{\pss{i}} \glex \str_j$. However, we also have $j \in (\pss{i}, i)$, which leads to the contradiction $\str_{\pss{i}} \llex \str_j$.
\end{proof}
\end{lemma}

\noindent Let $p_1 = i - 1, p_2, \dots, p_k = 0$ be the elements of $\pset_{i - 1}$ in descending order, then it follows from \cref{lemma:pssinclosure} that there is some $m \in [1, k]$ with $\pss{i} = p_m$, i.e.\ node $i$ has to become a child of $p_m$ in the partial PSS tree.
In terms of the BPS, we have to append $m - 1$ closing parentheses to the BPS prefix.
Then, we can simply write the opening parenthesis of node $i$.
Once again, an example is provided in \cref{fig:exampleproci} (right).

\begin{algorithm}
\caption{\nssconcept}
\label{alg:concept}
\newcommand{\ForInline}[2]{\State \algorithmicfor\ \ #1\ \ \algorithmicdo\ \ #2}
\begin{algorithmic}[1]
        \Require String $\str$ of length $n$
        \Ensure BPS of the PSS Tree of $\str$
        \State $\bpstree \gets ($ \Comment{Open node $0$}
        \For{$i = 1$ \textbf{to} $n$} \label{alg:concept:outerloop}
        \State Let $\pset_{i - 1} = \{p_1, \dots, p_k\}$ with $\pss{p_x} = p_{x + 1}$
        \State Determine $p_m = \pss{i}$ \label{alg:concept:determinem}
        \State Append $m - 1$ closing parentheses to $\bpstree$ \Comment{Close nodes $p_1, \dots, p_{m - 1}$}
        \State Append one opening parenthesis to $\bpstree$ \Comment{Open node $i$}
        \EndFor
        	\State Append $\absolute{\pset_{n}}$ closing parentheses to $\bpstree$ \Comment{Close rightmost path}
\end{algorithmic}
\end{algorithm}	
	
\noindent Our construction algorithm for $\pssbps$ is based on this simple idea, as outlined by \cref{alg:concept}. Initially, the BPS only contains the opening parenthesis of the root $0$ (line 1). Then, whenever we process an index $i$, we use $\pset_{i - 1}$ to determine $p_m$ (lines 3--4) and extend $\pssbps$ by appending $m - 1$ closing parentheses and one opening one (lines 5--6). Finally, once all nodes have been added to the PSS tree, we only have to close all remaining unclosed parentheses (line 7).

The algorithm has two black boxes: How do we determine $\pset_{i - 1}$ (line 3), and how do we use it to find $p_m$ (line 4)? The first question is easily answered, since the operations that we support on the BPS prefix at all times (see \cref{sec:auxmaintain}) are already sufficient to access each element of $\pset_{i - 1}$ in constant time. Let $p_1, \dots, p_k$ be exactly these elements in descending order. As explained earlier, they directly correspond to the unclosed parentheses of the BPS prefix, such that $p_k$ corresponds to the leftmost unclosed parenthesis, and $p_1$ to the rightmost one. Therefore, we have $p_x = \rankopen{\selectuncl{k - x + 1}} - 1$. It remains to be shown how to efficiently find $p_m$.

\subsection{Efficiently Computing \boldmath$p_m$\unboldmath}
\label{sec:proci}

Consider the following naive approach for computing $p_m$: Iterate over the indices $p_1, \dots, p_k$ in descending order (i.e.\ $p_1$ first, $p_k$ last). For each index $p_x$, evaluate whether $\str_{p_x} \llex \str_i$ holds. As soon as this is the case, we have found $p_m$. The cost of this approach is high: A naive suffix comparison between $\str_{p_x}$ and $\str_i$ takes $\lce{p_x, i} + 1$ individual character comparisons, which means that we spend ${\orderof{m + \sum_{x = 1}^m \lce{p_x, i}}}$ time to determine $m$. However, the following property will allow us to decrease this time bound significantly:
\begin{corollary}[Bitonic LCE Values]
	Let $p_1, \dots, p_k$ be exactly the elements of $\pset_{i - 1}$ in descending order and let $p_m = \pss{i}$. Furthermore, let $\ell_x = \lce{p_x, i}$ for all $x \in [1, k]$. We have $\ell_1 \leq \ell_2 \leq \dots \leq \ell_{m - 1}$ as well as $\ell_m \geq \ell_{m + 1} \geq \dots \geq \ell_k$.
	\begin{proof}
		Follows from $\str_{p_1} \glex \dots \glex \str_{p_{m - 1}} \glex \str_i \glex \str_{p_m} \glex \dots \glex \str_{p_k}$ and simple properties of the lexicographical order.
	\end{proof}
	\label{lemma:lcebitonic}
\end{corollary}

\begin{figure}
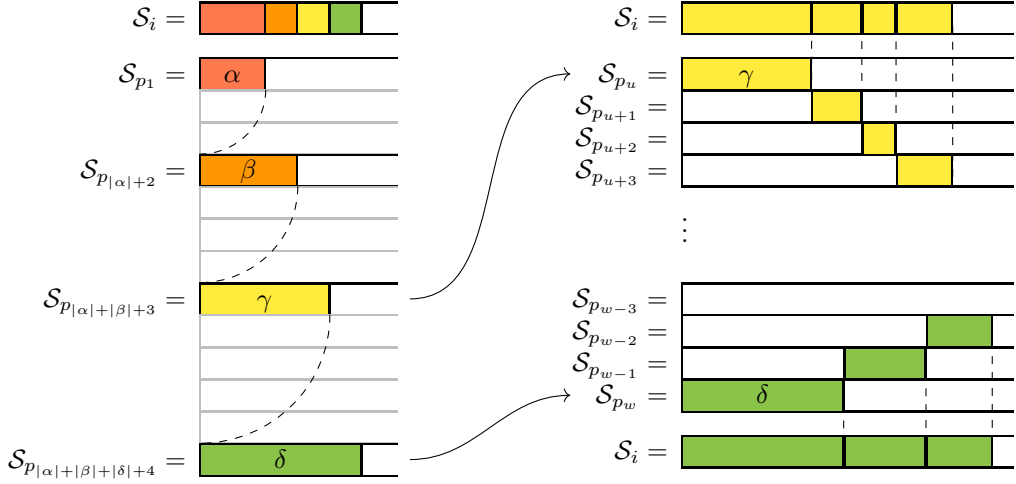

\small
\newlength{\singlecharspace}
\setlength{\singlecharspace}{.5\baselineskip}
\centering
\colorlet{alphacol}{highlight4}
\colorlet{betacol}{highlight3}
\colorlet{gammacol}{highlight1}
\colorlet{deltacol}{highlight2}
\colorlet{kappacol}{highlight1}
\newcommand{\isprefix}{\prec_{\textnormal{pref}}}
\renewcommand{\strboxphant}{\rule{1cm}{.375\baselineskip}}
\renewcommand{\newline}{\\[-.3\baselineskip]}
\newcommand{\anybox}[3][white]{\strboxhl{#1}{\hspace*{-\fboxsep}#2\mathclap{#3}#2\hspace*{-\fboxsep}}}
\newcommand{\spaceforbox}[1]{\hspace{#1\singlecharspace}}
\newcommand{\alphaboxempty}[1][]{\anybox[alphacol]{\spaceforbox2}{#1}}
\newcommand{\betaboxempty}[1][]{\anybox[betacol]{\spaceforbox3}{#1}}
\newcommand{\gammaboxempty}[1][]{\anybox[gammacol]{\spaceforbox3}{#1}}
\newcommand{\deltaboxempty}[1][]{\anybox[deltacol]{\spaceforbox4}{#1}}
\newcommand{\kappaboxempty}[1][]{\anybox[kappacol]{\spaceforbox5}{#1}}
\newcommand{\alphabox}{\tikzmark{l1}\alphaboxempty[\alpha]\tikzmark{r1}}
\newcommand{\betabox}{\tikzmark{l2}\betaboxempty[\beta]\tikzmark{r2}}
\newcommand{\gammabox}{\tikzmark{l3}\gammaboxempty[Z]\tikzmark{r3}}
\newcommand{\deltabox}{\tikzmark{l4}\deltaboxempty[\gamma]\tikzmark{r4}}
\newcommand{\kappabox}{\tikzmark{l5}\kappaboxempty[\delta]\tikzmark{r5}}
\newcommand{\finishline}{}
\newcommand{\openbox}{{\strlbox{\phantom{\kappabox}\quad}}}
\newcommand{\startline}{\rlap{\openbox}}
\newcommand{\dummyline}{&&&\color{gray!50!white}\startline\color{black}\newline}
\begin{minipage}[t]{.4\textwidth}
	\begin{alignat*}{4}
		\str_i = &&\enskip&%
		\rlap{\kappaboxempty}\rlap{\deltaboxempty}\rlap{\gammaboxempty}\rlap{\betaboxempty}\rlap{\alphaboxempty}\openbox\\[.3\baselineskip]
		\smash{\str_{p_{1}}} = &&&\startline\alphabox\tikzmark{udst}\finishline\newline
		\dummyline
		\dummyline
		\smash{\str_{p_{\absolute{\alpha} + 2}}} = &&&\startline\betabox\finishline\newline
		\dummyline
		\dummyline
		\dummyline
		\smash{\str_{p_{\absolute{\alpha} + \absolute{\beta} + 3}}} = &&&\startline\deltabox\tikzmark{usrc}\finishline\newline
		\dummyline
		\dummyline\tikzmark{wdst}
		\dummyline
		\dummyline
		\smash{\str_{p_{\absolute{\alpha} + \absolute{\beta} + \absolute{\delta} + 4}}} = &&&\startline\kappabox\tikzmark{wsrc}\finishline%
		\tikzol{
			\node (b1) at ($(l1)!0.6666!(l2)$) {};
			\node (b2) at ($(l2)!0.75!(l4)$) {};
			\node (b4) at ($(l4)!0.8!(l5)$) {};
			\foreach \x in {1,2,4} {
				\path (r\x.center) ++(0, -.2\baselineskip) node (r\x) {};
				\path (b\x.center) ++(0, -.2\baselineskip) node (b\x) {};
				\draw (r\x.center) edge[dashed, out=270, in=0] (b\x.center);
			};
			\node (srcx) at ($(usrc) + (3em, 0)$) {};
			\node (dstx) at ($(srcx) + (6em, 0)$) {};
			\node (usrc) at ($(usrc) + (0, .3em)$) {};
			\node (wsrc) at ($(wsrc) + (0, .3em)$) {};
			\node (udst) at ($(udst) + (0, .3em)$) {};
			\node (wdst) at ($(wdst) + (0, .3em)$) {};
			\draw (srcx |- usrc) edge[out=0, in=180, ->] (dstx |- udst);
			\draw (srcx |- wsrc) edge[out=0, in=180, ->] (dstx |- wdst);
		}%
	\end{alignat*}
	\end{minipage}~~~~~~~~~~~~~~~%
	\begin{minipage}[t]{.4\textwidth}
	\renewcommand{\openbox}{{\strlbox{\phantom{\kappabox}\qquad\qquad\qquad}}}
	\newcommand{\nextdelta}[1]{\anybox[deltacol]{\spaceforbox{#1}}{}}
	\newcommand{\nextkappa}[1]{\anybox[kappacol]{\spaceforbox{#1}}{}}
	\begin{alignat*}{4}
		\tikzmark{upperrow}\str_i = &&\enskip&%
		\rlap{\deltaboxempty\nextdelta{1.5}\nextdelta{1}\nextdelta{1.7}}%
		\openbox\\[.3\baselineskip]
		\smash{\str_{p_u} = }&&&\startline\deltabox\tikzmark{u1}\finishline\newline
		\smash{\str_{p_{u + 1}} = }&&&\startline\phantom{\deltabox}\nextdelta{1.5}\tikzmark{u2}\finishline\newline
		\smash{\str_{p_{u + 2}} = }&&&\startline\phantom{\deltabox\nextdelta{1.5}}\nextdelta{1}\tikzmark{u3}\finishline\newline
		\smash{\str_{p_{u + 3}} = }&&&\startline\phantom{\deltabox\nextdelta{1.5}\nextdelta{1}}\nextdelta{1.7}\tikzmark{u4}\finishline\newline
		&&&\vphantom{\startline}\smash{}\newline
		&&&\vphantom{\startline}\smash{\vdots}\newline
		&&&\vphantom{\startline}\smash{}\newline
		\smash{\str_{p_{w - 3}} = }&&&\startline\finishline\newline
		\smash{\str_{p_{w - 2}} = }&&&\startline\phantom{\kappabox\nextkappa{2.5}}\nextkappa{2}\tikzmark{w3}\finishline\newline
		\smash{\str_{p_{w - 1}} = }&&&\startline\phantom{\kappabox}\nextkappa{2.5}\tikzmark{w2}\finishline\newline
		\smash{\str_{p_w} = }&&&\startline\kappabox\tikzmark{w1}\finishline\\[.3\baselineskip]%
		\tikzmark{lowerrow}\str_i = &&&%
		\rlap{\kappaboxempty\nextkappa{2.5}\nextkappa{2}}%
		\openbox%
		\tikzol{
			\foreach \x in {1,2,3,4} {
				\draw (u\x) edge[dashed] (u\x |- upperrow);
			};
			\foreach \x in {1,2,3} {
				\draw (w\x) edge[dashed] (w\x |- lowerrow);
			};
		}%
	\end{alignat*}
	\end{minipage}
	\vspace{.5\baselineskip}
	
	\caption{Matching character comparisons when determining $p_m$. On the left we have the suffix $\str_i$ as well as $\str_{p_1}, \str_{p_2}, \dots, \str_{p_w}$, which are relevant for the first step. Each prefix $\alpha, \beta, \gamma, \delta$ highlights the LCP between the respective suffix $\str_{p_x}$ and $\str_i$. On the right side we have the suffixes $\str_{p_u}, \str_{p_{u + 1}}, \dots, \str_{p_w}$, which are relevant for the second step.}
	\label{fig:findm}
\end{figure}

\noindent From now on, we continue using the notation $\ell_x = \lce{p_x, i}$ from the corollary. Note that the longest LCE between $i$ and any of the $p_x$ occurs either with $p_m$ or with $p_{m - 1}$. Let $\ellmax = \max(\ell_{m - 1}, \ell_m)$ be this largest LCE value, then our more sophisticated approach for determining $m$ only takes ${\orderof{m + \ellmax}}$ time. It consists of two steps: First, we determine a candidate interval $(u, w] \subseteq [1, k]$ of size at most $\ellmax$ that contains $m$. 
In the second step we gradually narrow down the borders of the candidate interval until the exact value of $m$ is known.
\paragraph{Step 1: Find a candidate interval.} Our goal is to find ${(u, w] = (u, u + \ell_u + 1]}$ with $m \in (u, w]$. Initially, we naively compute $\ell_1 = \lce{p_1, i}$, allowing us to evaluate $\str_{p_1} \llex \str_i$ in constant time. If this holds, then we have $m = 1$ and no further steps are necessary. Otherwise, let $u \gets 1$ and (i) let $\unew \gets u + \ell_u + 1$. We already know that $u < m$ holds. Now we have to evaluate if $m \leq w$ also holds. Therefore, we compute $\ell_\unew = \lce{p_\unew, i}$ naively, which allows us to check in constant time if $\str_{p_\unew}\llex \str_i$ and decide if $m \leq w$ holds. If this is not the case, then we assign $u \gets \unew$ as well as $\ell_u \gets \ell_\unew$ and continue at~(i). If however $\str_{p_\unew} \llex \str_i$ holds, then we have $m \leq \unew$ and therefore $m \in (u, \unew]$. \cref{fig:findm} (left) outlines the procedure.

\paragraph{Step 2: Narrow down \boldmath$(u, w]$ to the exact value of $m$.\unboldmath} Now we gradually tighten the borders of the candidate interval. If $\ell_u$ is smaller than $\ell_w$, then we try to increase $u$ by one. Otherwise, we try to decrease $w$ by one. 

Assume that we have $\ell_u < \ell_w$, then it follows from \cref{lemma:lcebitonic} that $\ell_{u + 1} \geq \ell_u$ holds.
Therefore, when computing $\ell_{u + 1}$ we can simply skip the first $\ell_u$ character comparisons.
Now we use $\ell_{u + 1}$ to evaluate in constant time if $\str_{p_{u + 1}} \glex \str_{i}$ holds.
If that is the case, then we have $u + 1 < m$ and thus we can assign $u \gets u + 1$ and start Step 2 from the beginning.
If however $\str_{p_{u + 1}} \llex \str_{i}$ holds, then we have $m = u + 1$ and no further steps are necessary.
In case of $\ell_u \geq \ell_w$ we proceed analogously. Once again, \cref{fig:findm} (right) visualizes the procedure.

\paragraph{Time Complexity.}
Step 1 is dominated by computing LCE values.
Determining the final LCE value $\ell_w$ takes $\ell_w + 1$ individual character comparisons and thus $\thetaof{\ell_w + 1}$ time.
Whenever we compute any previous value of $\ell_w$, we increase $w$ by $\ell_w + 1$ afterwards.
Therefore, the time for computing all LCE values is bound by $\thetaof{w + \ell_w} = \thetaof{u + \ell_u + \ell_w} \subseteq \orderof{m + \ellmax}$.

Since initially $(u, w]$ has size at most $\ellmax$, we call Step 2 at most $\orderof{\ellmax}$ times.
With every call we increase $\ell_u$ or $\ell_w$ by exactly the number of matching character comparisons that we perform.
Therefore, the total number of matching character comparisons is bound by $2\ellmax$.
Thus, the total time needed for Step 2 is bound by $\orderof{\ellmax}$.

In sum, processing index $i$ takes $\orderof{m + \ellmax}$ time.
For the total processing time of all indices (and thus the execution time of \cref{alg:concept}) we get:
\begin{alignat*}{8}
    &\sum_{i = 1}^n &\ &\orderof{\ \overbrace{\absolute{\pset_{i - 1} \cap [\pss{i}, i]}}^m\ } &\ +\ & \sum_{i = 1}^n &\ &\orderof{\ {\overbrace{{\max}_{p_x \in \pset_{i - 1}} \lce{p_x, i}}^{\ellmax}}\ }\\
    = &&&\orderof{n} &\ +\ &&&\orderof{n^2}
\end{alignat*}
(The $\orderof{m}$-terms sum to $n$ since $m - 1$ is exactly the number of closing parentheses that we write while processing $i$, and there are exactly $n + 1$ closing parentheses in the entire BPS.)
As it appears, the total time bound of the algorithm is still far from linear time.
However, it is easy to identify the crucial time component that makes the algorithm too expensive.
From now on we call the $\orderof{m}$ term of the processing time \emph{negligible}, while the $\orderof{\ellmax}$ term is called \emph{critical}. 

Clearly, if we could somehow remove the critical terms, we would already achieve linear time.
In the following section we propose a technique of amortizing the critical terms such that on average the critical term per index becomes constant.
This way, the execution time of \cref{alg:concept} decreases to $\orderof{n}$.

\section{Achieving Linear Time}

The critical time component for processing index $i$ is $\ellmax = {\max}_{p_x \in \pset_{i - 1}} \lce{p_x, i}$. When processing $i$ with the technique from \cref{sec:proci}, we inherently find out the exact value of $\ellmax$, and we also discover the index $\pmax$ for which we have $\lce{\pmax, i} = \ellmax$. From now on, we %
simply use $\ell = \ellmax$ and $j = \pmax$. 
While discovering a large LCE value $\ell$ is costly, it yields valuable structural information about the input text: There is a repeating substring of length $\ell$ with occurences $\str[j..j+\ell)$ and $\str[i..i+\ell)$. Intuitively, there is also a large repeating structure in the PSS tree, and consequently a repeating substring in $\bpstree$. This motivates the techniques shown in this section, which conceptually alter \cref{alg:concept} as follows: Whenever we finish processing an index $i$ with critical cost $\ell$, we skip the next $\Omega(\ell)$ iterations of the loop by simply extending the BPS prefix with the copy of an already computed part, which means that the amortized critical cost per index becomes constant. 

Depending on $j$ and $\ell$ we choose either the \emph{run extension} (\cref{sec:runextend}) or the \emph{amortized look-ahead} (\cref{sec:amolook}) to perform the extension. Before going into detail, we point out that $\str[j..i)$ is a Lyndon word. As mentioned earlier, it follows from \cref{lemma:lcebitonic} that $j$ equals $p_m$ or $p_{m-1}$. Since $i$ is the first node that is not a descendant of $p_{m - 1}$, we have $\nss{p_{m - 1}} = i$. Therefore, if $j = p_{m - 1}$ holds, we have $\nss{j} = i$, which by definition implies that $\str[j..i)$ is a Lyndon word. If however $j = p_m = \pss{i}$ holds, then $\str[j..i)$ is a Lyndon word because of \cref{lemma:psslyndonword}.

\subsection{Run Extension}

\label{sec:runextend}
\newcommand{\alphamax}{\ensuremath{{\gamma}}}

We apply run extension iff we have $\ell \geq 2(i - j)$. It is easy to see that in this case $\str[j..j+\ell)$ and $\str[i..i+\ell)$ overlap such that the Lyndon word $\rungreek = \str[j..i)$ repeats itself at least three times, starting at index $j$. We call the substring $\str[j..i + \ell)$ \emph{Lyndon run with period $\absolute{\rungreek}$}. The number of \emph{repetitions} is $t = \floor{\ell / \absolute{\rungreek}} + 1 \geq 3$, and the starting positions of the repetitions are $r_1, \dots, r_t$ with $r_1 = j$, $r_2 = i$, and generally $r_x = r_{x - 1} + \absolute{\rungreek}$. 
In a moment we will show that in this particular situation the following lemma holds:

\begin{lemma}
Let $o_x$ be the index of the opening parenthesis of node $x$ in $\pssbps$. Then we have $\pssbps[o_{r_1}..o_{r_2}] = \pssbps[o_{r_2}..o_{r_3}] = \dots = \pssbps[o_{r_{t - 1}}..o_{r_t}]$.
\label{lemma:REmain}
\end{lemma}

Expressed less formally, each repetition of $\rungreek$ --- except for the last one --- induces the same substring in the BPS. Performing the run extension is as easy as taking the already written substring $\pssbps(o_{r_1}..o_{r_2}] = \pssbps(o_{j}..o_{i}]$, and appending it $t - 2$ times to $\pssbps$. Afterwards, the last parenthesis that we have written is the opening parenthesis of node $r_t$, and we continue the execution of \cref{alg:concept} with iteration $r_t + 1$. Thus, we have skipped the processing of $r_t - i$ indices. Since 
\begin{alignat*}{1}
    r_t - i\ =\ (t - 2) \cdot \absolute{\rungreek} \ \geq\ \frac{(t - 2) \cdot \absolute{\rungreek}}{t \cdot \absolute{\rungreek}} \cdot \ell \ \geq\ \frac{1}{3} \cdot \ell \ =\ \Omega(\ell)\ ,
\end{alignat*}
it follows that the average critical cost per index from $[i, r_t]$ is constant.

\label{sec:REproof}

\paragraph{Proving the Lemma.} It remains to be shown that \cref{lemma:REmain} holds. It is sufficient to prove the correctness for $t = 3$, since the correctness for the general case follows by repeatedly applying the lemma with $t = 3$. Therefore, we only have to show $\pssbps[o_{r_1}..o_{r_2}] = \pssbps[o_{r_2}..o_{r_3}]$.

\paragraph{Isomorphic Subtrees.}%
%
Since $\rungreek$ is a Lyndon word, it is easy to see that the suffixes at the starting positions of repetitions are lexicographically smaller than the suffixes that begin in between the starting positions of repetitions, i.e.\ we have $\forall x \in (r_1, r_2) : \str_{r_1} \llex \str_x$ and $\forall x \in (r_2, r_3) : \str_{r_2} \llex \str_x$. Consequently, the indices from $(r_1, r_2)$ are descendants of $r_1$ in the PSS tree, and the indices from $(r_2, r_3)$ are descendants of $r_2$ in the PSS tree, i.e.\ each of the intervals $[r_1, r_2)$ and $[r_2, r_3)$ induces a tree. 

Next, we show that these trees are actually isomorphic. Clearly, the tree induced by $[r_1, r_2)$ solely depends on the lexicographical order of suffixes that begin within $[r_1, r_2)$, and the tree induced by $[r_2, r_3)$ solely depends on the lexicographical order of suffixes that begin within $[r_2, r_3)$. Assume that the trees are \emph{not} isomorphic, then there must be a suffix comparison that yields different results in each interval, i.e.\ there must be offsets $a, b \in [0, \absolute{\rungreek})$ with $a \neq b$ such that $\str_{r_1 + a} \llex \str_{r_1 + b} \Longleftrightarrow \str_{r_2 + a} \glex \str_{r_2 + b}$ holds. However, this is impossible, as shown by the lemma below.

%
\begin{lemma}\label{lemma:compareinrun}\label{lemma:compareinrun2}
    For all $a, b \in [0, \absolute{\rungreek})$ with $a \neq b$ we have $\str_{r_1 + a} \llex \str_{r_1 + b} \Longleftrightarrow \str_{r_2 + a} \llex \str_{r_2 + b}$.
    \begin{proof}
        Assume w.l.o.g.\ $a < b$, and let $a' = a + 1$ and $b' = b + 1$. We can show that the strings $\rungreek_{a'}\cdot\rungreek$ and $\rungreek_{b'}\cdot\rungreek$ have a mismatch:\\
        \newcommand{\anothermu}{{\strbox{\qquad\qquad\rungreek\qquad\qquad}}}%
        \newcommand{\musub}[1]{\mathrlap{\rungreek_{#1}}\phantom{\rungreek}}%
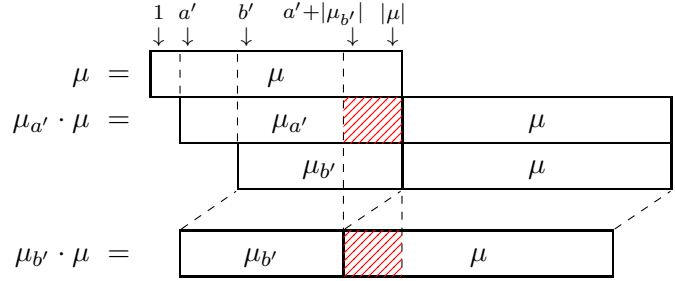
        \begin{alignat*}{10}
            \rungreek\ =\ &\ &\strlbox{\stridxarrow{1}\quad}&\strmbox{\stridxarrow{a'}\qquad}&&\strmbox{\stridxarrow{b'}\quad\mathclap{\rungreek}}&&\strmbox{\qquad}&&\strrbox{\mathrlap{\phantom{\strbox{}}\stridxarrow{a'+\absolute{\rungreek_{b'}}\qquad}}\qquad\stridxarrow{\absolute{\rungreek}}}
            \\[-.425em]
            \rungreek_{a'}\cdot\rungreek\ =\ &&\strlbox{}&\strmbox{\qquad}&&\strmbox{\quad\mathclap{\quad\rungreek_{a'}}}&&\strmbox{\qquad}&&\strrbox{\qquad}\anothermu\\[-.425em]
            &&&&&\mathllap{\strlbox{}}\strmbox{\quad\mathclap{\qquad\quad\rungreek_{b'}}}&&\strmbox{\qquad}&&\strrbox{\qquad}\anothermu\\[1em]
            \rungreek_{b'} \cdot \rungreek\ =\ &&\strlbox{}&\strmbox{\qquad}&&\strmbox{\enskip\mathclap{\rungreek_{b'}}\enskip}&&\strmbox{\qquad}&&\strrbox{}\anothermu\\[-1.45em]
            &&\tikzmark{ll1}\hspace{\fboxsep}&&&\hspace{-\fboxsep}\hspace{-\fboxrule}\tikzmark{ll2}&&&&\phantom{\strrbox{}}\tikzmark{ll3}\qquad\tikzmark{ll4}\hphantom{\anothermu}\mkern-36mu\tikzmark{ll5}\qquad\tikzmark{ll6}%
            \begin{tikzpicture}[overlay, remember picture, every node/.style={inner sep=0pt, outer sep=0pt}]
            \foreach \x in {3,4} {
                \draw[dashed] (ll\x) ++(0, 2.975cm) -- (ll\x);
            }
            \foreach \x in {1,2} {
                \path (ll\x) ++(0, 2.975cm) node[inner sep=0pt, outer sep=0pt] (lln\x) {};
                \draw[dashed] (ll\x) ++(0, 1.8cm) -- (lln\x);
            }
            \path (ll1) ++(0, 0.625cm) node[inner sep=0pt, outer sep=0pt] (lln1) {};
            \path (ll2) ++(0, 1.15cm) node[inner sep=0pt, outer sep=0pt] (lln2) {};
            \path (ll3) ++(0, 0.625cm) node[inner sep=0pt, outer sep=0pt] (lln3) {};
            \path (ll4) ++(0, 1.15cm) node[inner sep=0pt, outer sep=0pt] (lln4) {};
            \path (ll5) ++(0, 0.625cm) node[inner sep=0pt, outer sep=0pt] (lln5) {};
            \path (ll6) ++(0, 1.15cm) node[inner sep=0pt, outer sep=0pt] (lln6) {};
            \draw[dashed] (lln1) -- (lln2) (lln3) -- (lln4) (lln5) -- (lln6);
            \path (ll3) ++(0, 0.005cm) node[inner sep=0pt, outer sep=0pt] (r1) {};
            \path (ll4) ++(0, 0.625cm) node[inner sep=0pt, outer sep=0pt] (r2) {};
            \path (r1) ++(0, 1.765cm) node[inner sep=0pt, outer sep=0pt] (r3) {};
            \path (r2) ++(0, 1.765cm) node[inner sep=0pt, outer sep=0pt] (r4) {};
            \fill[pattern=north east lines, pattern color=red, dashed] (r1) rectangle (r2) (r3) rectangle (r4);
            \end{tikzpicture}\\[-2.5em]
        \end{alignat*}
        Consider the two hatched areas in the drawing above. The top area highlights the suffix $\rungreek_{a' + \absolute{\rungreek_{b'}}}$ of $\rungreek$, which has length $c = \absolute{\rungreek} - (a' + \absolute{\rungreek_{b'}}) + 1
        $. The bottom area highlights the prefix $\rungreek[1..c]$ of $\rungreek$. Since $\rungreek$ is a Lyndon word, there is no proper non-empty suffix of $\rungreek$ that is also a prefix of $\rungreek$. It follows that the hatched areas cannot be equal, i.e.\ $\rungreek_{a' + \absolute{\rungreek_{b'}}} \neq \rungreek[1..c]$. This guarantees a mismatch between $\rungreek_{a'}\cdot\rungreek$ and $\rungreek_{b'}\cdot\rungreek$. Therefore, appending an arbitrary string to $\rungreek_{a'}\cdot\rungreek$ and $\rungreek_{b'}\cdot\rungreek$ does not influence the outcome of a lexicographical comparison. The statement of the lemma directly follows by appending $\str_{r_3}$ and $\str_{r_4}$ respectively:
        \begin{alignat*}{4}
            \rungreek_{a'}\cdot\rungreek \llex \rungreek_{b'}\cdot\rungreek\ \ \Longleftrightarrow\ \ &&\underbrace{\rungreek_{a'}\cdot\rungreek\cdot\str_{r_{3}}}_{=\ \str_{r_1 + a}}\ \ \llex\ \ & \underbrace{\rungreek_{b'}\cdot\rungreek\cdot\str_{r_{3}}}_{=\ \str_{r_1 + b}}\ \ \Longleftrightarrow\ \ &&\underbrace{\rungreek_{a'}\cdot\rungreek\cdot\str_{r_{4}}}_{=\ \str_{r_2 + a}}\ \ \llex\ \ & \underbrace{\rungreek_{b'}\cdot\rungreek\cdot\str_{r_{4}}}_{=\ \str_{r_2 + b}}
        \end{alignat*}
        
        {\color{white}.\\[-1.5\baselineskip]}
    \end{proof}
\end{lemma}

Finally, we show that in the PSS tree the induced isomorphic trees are connected in a way that ultimately implies $\pssbps[o_{r_1}..o_{r_2}] = \pssbps[o_{r_2}..o_{r_3}]$. There are two possible scenarios for this connection, which depend on the so called \emph{direction} of the Lyndon run. We call a run \emph{increasing} iff $\str_{r_1} \llex \str_{r_2}$ holds, and \emph{decreasing} otherwise.

\subsubsection*{Increasing Runs}
\label{sec:incrun}

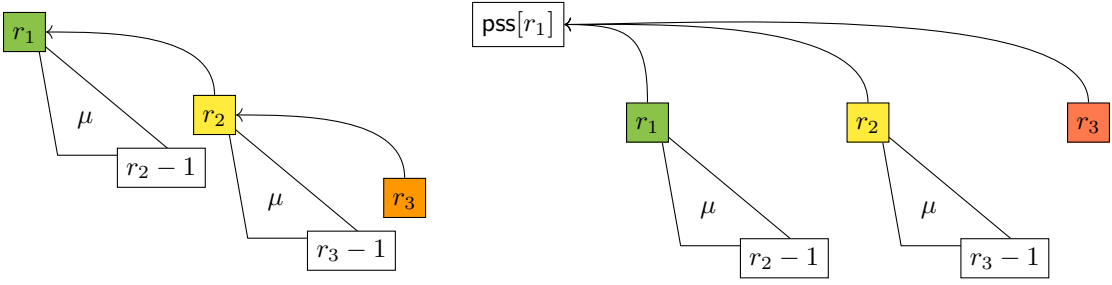
\begin{figure}
	\small
    \newcommand{\maybehlseven}{white}%
    \centering
    \tikzset{
        periodtree/.style={
            draw,shape border uses incircle,
            isosceles triangle,isosceles triangle apex angle=100,
            shape border rotate=230,minimum height=1cm,fill=\maybehlseven}
    }
    \newcommand{\periodtreelabel}{\ensuremath{\smash{\mathclap{\rungreek}}}}
    \newcommand{\periodtreehscale}{1}
    \newcommand{\periodtreevscale}{1}
    \newcommand{\mynode}[1]{$\vphantom{i}#1$}

    \newcommand{\fcb}[2]{%
        \setlength\defaultfboxsep{\fboxsep}%
        \setlength\fboxsep{0pt}%
        \fbox{%
            \setlength\fboxsep{\defaultfboxsep}%
            \colorbox{#1}{#2}}%
        \setlength\fboxsep{\defaultfboxsep}}

    \begin{tikzpicture}
    \draw[]
    node[periodtree]{\periodtreelabel}
    ++(-.8cm, 1.2cm)
    node[draw, fill=highlight1] (r1) {\mynode{r_1}}
    ++(1.8cm, -1.8cm)
    node[draw, fill=\maybehlseven]{\mynode{r_2 - 1}}
    ++(1.5cm, -.5cm)
    node[periodtree]{\periodtreelabel}
    ++(-.8cm, 1.2cm)
    node[draw, fill=highlight2] (r2) {\mynode{r_2}}
    ++(1.8cm, -1.8cm)
    node[draw, fill=\maybehlseven]{\mynode{r_3 - 1}}
    ++(1.5cm, -.5cm)
    ++(-.8cm, 1.2cm)
    node[draw, fill=highlight3] (rt) {\mynode{r_{3}}};
    \draw (r2.north) edge[out=90, in=0, looseness=.9, ->] (r1.east) (rt.north) edge[out=90, in=0, looseness=.9, ->] (r2.east) ;
    \end{tikzpicture}~%
\begin{tikzpicture}
    \draw[]
    node[draw, fill=white] (p) {\mynode{\pss{r_1}}}
    ++(2.5cm, -2.5cm)
    node[periodtree]{\periodtreelabel}
    ++(-.8cm, 1.2cm)
    node[draw, fill=highlight1] (r1) {\mynode{r_1}}
    ++(1.8cm, -1.8cm)
    node[draw, fill=\maybehlseven]{\mynode{r_2 - 1}}
    ++(0, 1.8cm)
    ++(1.9cm, -1.2cm)
    node[periodtree] (triangle1) {\periodtreelabel}
    ++(-.8cm, 1.2cm)
    node[draw, fill=highlight2] (r2) {\mynode{r_2}}
    ++(1.8cm, -1.8cm)
    node[draw, fill=\maybehlseven] (r3m) {\mynode{r_3 - 1}}
    ++(0cm, 1.8cm)
    ++(1.9cm, -1.2cm)
    ++(-.8cm, 1.2cm)
    node[draw, fill=highlight4] (rt) {\mynode{r_{3}}}
    ++(0, -2.2cm);
    \draw (r1.north) edge[out=90, in=0, looseness=1.5, ->] (p.east);
    \draw (r2.north) edge[out=90, in=0, looseness=.7, ->] (p.east);
    \draw (rt.north) edge[out=90, in=0, looseness=.5, ->] (p.east);
    \end{tikzpicture}    
    \caption{Connections between run-induced subtrees in the PSS tree, depending on the direction of the run (left: increasing run, right: decreasing run).}
    \label{fig:isotrees}
\end{figure}

First, we focus on increasing runs. It follows from $\str_{r_1} \llex \str_{r_2}\ \Longleftrightarrow\ \rungreek\cdot\str_{r_2} \llex \rungreek\cdot\str_{r_3} \Longleftrightarrow \str_{r_2} \llex \str_{r_3}$ that $\str_{r_1} \llex \str_{r_2} \llex \str_{r_3}$. Since $\rungreek$ is a Lyndon word, we have $\forall x \in (r_1, r_2) : \str_{r_2} \llex \str_x$ as well as $\forall x \in (r_2, r_3) : \str_{r_3} \llex \str_x$. Therefore, we have $\pss{r_2} = r_1$ and $\pss{r_3} = r_2$, and the isomorphic subtrees are connected as visualized in \cref{fig:isotrees} (left).
Therefore we have $\pssbps[o_{r_1}..o_{r_2}] = \pssbps[o_{r_2}..o_{r_3}]$, which means that \cref{lemma:REmain} holds for increasing runs.

\subsubsection*{Decreasing Runs}

With the same argument as for increasing runs, we have $\str_{r_1} \glex \str_{r_2} \glex \str_{r_3}$ in decreasing runs.
We also have $\forall x \in (r_1, r_2) : \str_{r_2} \llex \str_x$ as well as $\forall x \in (r_2, r_3) : \str_{r_3} \llex \str_x$, which means that $\pss{r_2} \leq \pss{r_1}$ and $\pss{r_3} \leq \pss{r_1}$ hold. In \cref{lemma:REdecpss} we will show that in fact $\pss{r_1} = \pss{r_2} = \pss{r_3}$ holds, such that the isomorphic subtrees are connected as visualized in \cref{fig:isotrees} (right). 
Therefore we have $\pssbps[o_{r_1}..o_{r_2}] = \pssbps[o_{r_2}..o_{r_3}]$, which means that \cref{lemma:REmain} holds for decreasing runs.

\Needspace{7\baselineskip}\begin{lemma}
    In decreasing runs we have $\pss{r_1} = \pss{r_2} = \pss{r_3}$.
    \begin{proof}
    	As explained previously, we have $\pss{r_2} \leq \pss{r_1}$ and $\pss{r_3} \leq \pss{r_1}$, and thus only need to show $\str_{\pss{r_1}} \llex \str_{r_2}$ and $\str_{\pss{r_1}} \llex \str_{r_3}$.
        We will show below that $\rungreek$ cannot a prefix of $\str_{\pss{r_1}}$, from which the statement of the lemma can be deduced easily
        since the suffixes $\str_{r_2}$ and $\str_{r_3}$ begin with the prefix $\rungreek$.

        Assume for the sake of contradiction that $\rungreek$ is a prefix of $\str_{\pss{r_1}}$. If we also assume
        $\pss{r_1} + \absolute{\rungreek} > r_1$, we get:\\
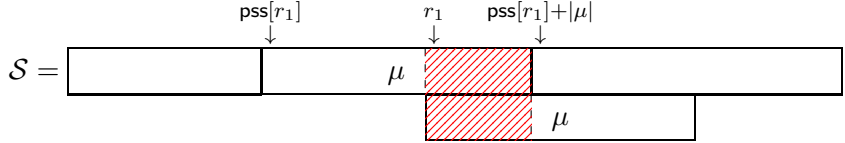
        \begin{alignat*}{2}
            \str = \strbox{\qquad\qquad\qquad}%
            \strlbox{\stridxarrow{\pss{r_1}}\qquad\qquad\rungreek\quad}%
            &\strrbox{\stridxarrow{r_1}\quad\qquad}%
            \tikzmark{rr}%
            \strbox{\stridxarrow{\pss{r_1} + \absolute{\rungreek}}\qquad\qquad\qquad\qquad\qquad}\\[-.425em]
            \tikzmark{ll}\strlbox{}%
            &\mathrlap{\strrbox{\qquad\qquad\rungreek\qquad\qquad}}%
            \begin{tikzpicture}[overlay, remember picture, every node/.style={inner sep=0pt, outer sep=0pt}]
                \path (rr |- ll) node (1) {};
                \path (ll |- rr) node (2) {};
                \foreach \x in {1,2} {
                    \path (\x) ++(0, -.55em) node (\x) {};
                    \draw[dashed] (\x) ++(0, 1.6em) node (up\x) {} -- (\x);
                }
                \fill[pattern=north east lines, pattern color=red, dashed] (up2) rectangle (1);
            \end{tikzpicture}
        \end{alignat*}
        As indicated by the hatched area, this implies that there is a proper non-empty suffix of $\rungreek$ that is also a prefix of $\rungreek$, which is impossible because $\rungreek$ is a Lyndon word. Thus we have $\pss{r_1} + \absolute{\rungreek} \ngtr r_1$. Also, we cannot have $\pss{r_1} + \absolute{\rungreek} = r_1$, because then $\pss{r_1}$ would be the starting position of another repetition of $\rungreek$, which would imply $\str_{\pss{r_1}} \glex \str_{r_1}$.
        %
        %
        It follows $\pss{r_1} + \absolute{\rungreek} < r_1$, i.e.\ ${\pss{r_1} + \absolute{\rungreek}} \in (\pss{r_1}, r_1)$ and thus $\str_{\pss{r_1} + \absolute{\rungreek}} \glex \str_{r_1}$. However, this leads to a contradiction:%
        \begin{alignat*}{4}
        \str_{\pss{r_1}} \llex \str_{r_1}\  \Longleftrightarrow\ &\ &\rungreek \cdot \str_{\pss{r_1} + \absolute{\rungreek}} &\llex \rungreek \cdot \str_{r_2}\\
        \Longleftrightarrow\ &&\str_{\pss{r_1} + \absolute{\rungreek}} &\llex \str_{r_2}\\
        \underset{\vphantom{I^I}\mathclap{\str_{r_1} \glex \str_{r_2}}}{\Longrightarrow}\ &&\str_{\pss{r_1} + \absolute{\rungreek}} &\llex \str_{r_1}
        \end{alignat*}
    \end{proof}
\label{lemma:REdecpss}
\end{lemma}

\subsection{Amortized Look-Ahead}
\label{sec:amolook}

Finally, we show how to amortize the critical cost $\orderof{\ell}$ of processing index $i$ if the run extension is not applicable, i.e.\ if we have $\ell < 2(i - j)$. Unfortunately, the trees induced by the nodes from $[j, j + \ell)$ and $[i, i + \ell)$ are not necessarily isomorphic. However, we can still identify a sufficiently large isomorphic structure. In a moment we will show that the following lemma holds:

\begin{lemma}
Let $o_x$ be the index of the opening parenthesis of node $x$ in $\pssbps$. 
We either have $\pssbps[o_j..o_{j + \floor{\ell/4} - 1}] = \pssbps[o_i..o_{i + \floor{\ell/4} - 1}]$, or there is an integer 
$\anchor < \floor{\ell/4}$ with $\pssbps[o_j..o_{j + \anchor - 1}] = \pssbps[o_i..o_{i + \anchor - 1}]$ and 
an index $h \in [i, i + \anchor)$ such that $\str[h..i + \ell)$ is a Lyndon run of the Lyndon word $\str[h..i + \anchor)$. We can determine which case applies, and also determine the value of $\anchor$ (if applicable) in $\orderof{\ell}$ time.
\label{lemma:ALmain}
\end{lemma}

When performing the amortized look-ahead we first determine which case of the lemma applies.
Then, if $\pssbps[o_j..o_{j + \floor{\ell/4} - 1}] = \pssbps[o_i..o_{i + \floor{\ell/4} - 1}]$, we extend the known prefix of the BPS by appending a copy of $\pssbps[o_j..o_{j + \floor{\ell/4} - 1}]$, and continue the execution  of \cref{alg:concept} with iteration $i + \floor{\ell/4}$.
Since this way we skip the processing of $\floor{\ell/4} - 1 = \Omega(\ell)$ indices, the average critical cost per index from $[i, i + \floor{\ell/4})$ is constant.

If, however, the second case applies, then we determine the value of $\anchor$ and extend the known prefix of the BPS by appending a copy of $\pssbps[o_j..o_{j + \anchor - 1}]$, allowing us to continue the execution  of \cref{alg:concept} with iteration $i + \anchor$.
We know that there is some $h \in [i, i+\anchor)$ such that $\str[h..i + \ell)$ is a Lyndon run of the Lyndon word $\rungreek = \str[h..i + \anchor)$.
This run might even be longer: Let $\ell' = \lce{h, i + \anchor}$ (computed naively), then $\str[h..i + \anchor + \ell')$ is the longest run of $\rungreek$ that starts at index $h$.
If the run is increasing, then $\pss{i + \anchor} = h$ holds (see \cref{sec:incrun}), and the longest LCE that we discover when processing index $i + \anchor$ is $\ell'$.
If the run is decreasing, then $\pss{i + \anchor} = \pss{h}$ holds.
Also in this case, the longest LCE that we discover when processing index $i + \anchor$ is $\ell'$, since $\lce{\pss{i + \anchor}, i + \anchor}$ is less than $\absolute{\rungreek}$ (see proof of \cref{lemma:REdecpss}).
Therefore, the critical cost of processing index $i + \anchor$ is $\orderof{\ell'}$.
However, since the Lyndon run has at least three repetitions, we will also skip the processing of $\Omega(\ell')$ indices by using the run extension.

The algorithmic procedure for the second case can be summarized as follows: We process index $i$ with critical cost $\orderof{\ell}$ and skip $\anchor - 1$ indices afterwards. Then we process index $i + \anchor$ with critical cost $\orderof{\ell'}$ and skip another $\Omega(\ell')$ indices by using the run extension. Since we have $\ell' = \Omega(\ell)$, the total critical cost is $\orderof{\ell'}$, and the total number of processed or skipped indices is $\Omega(\ell')$. Thus, the average critical cost per index is constant.
\vspace{.5\baselineskip}

\paragraph{Proving the Lemma.} It remains to be shown that \cref{lemma:ALmain} holds. For this purpose, assume $\pssbps[o_j..o_{j + \floor{\ell/4} - 1}] \neq \pssbps[o_i..o_{i + \floor{\ell/4} - 1}]$.
From now on we refer to $\pssbps[o_j..o_{j + \floor{\ell/4} - 1}]$ and $\pssbps[o_i..o_{i + \floor{\ell/4} - 1}]$ as \emph{left} and \emph{right side}, respectively.
Consider the first mistmatch between the two, where w.l.o.g.\ we assume that the mismatch has an opening parenthesis on the left side, and a closing one on the right side.
On the left side, the opening parenthesis corresponds to a node $j + x$ with $x \in [1, \floor{\ell/4})$ that is a child of another node $j + h$.
Since $\str[j..j + \ell)$ is a Lyndon word, all nodes from $(j, j + \ell)$ are descendants of $j$.
Consequently, we have $h \in [0, x)$. Now we look at the right side: Since we have a closing parenthesis instead of an opening one, we know that $i + x$ is not attached to $i + h$, but to a smaller node, i.e.\ we have $\pss{i + x} < i + h$.
It follows that $\str_{j + h} \llex \str_{j + x}$ and $\str_{i + h} \glex \str_{i + x}$ hold.
Since these suffix comparisons yield different results, but at the same time we have $\str[j..j + \ell) = \str[i..i + \ell)$, it follows that the mismatch between $\str_{j + h}$ and $\str_{j + x}$ (respectively between $\str_{i + h}$ and $\str_{i + x}$) cannot lie within $\str[j..j + \ell)$ (respectively $\str[i..i + \ell)$), and thus we have $\lce{j + h, j + x} > \ell - x$.

Now we show that $\str[j + h..j + \ell)$ is a Lyndon run with period $x - h$.
Since $\pss{j + x} = j + h$ holds, it follows from \cref{lemma:psslyndonword} that $\str[j + h..j + x)$ is a Lyndon word.
Due to $\lce{j + h, j + x} > \ell - x \geq 3(\ell/4) \geq 3(x - h)$ we know that the Lyndon word repeats at least four times, and the run extends all the way to the end of $\str[j..j + \ell)$.
Note that since the opening parenthesis of node $j + x$ causes the first mismatch between $\pssbps[o_j..o_{j + \floor{\ell/4} - 1}]$ and $\pssbps[o_i..o_{i + \floor{\ell/4} - 1}]$, we have $\pssbps[o_j..o_{j + x - 1}] = \pssbps[o_i..o_{i + x - 1}]$.
Therefore, $\anchor \gets x$ already satisfies \cref{lemma:ALmain}.
\vspace{.5\baselineskip}

Finally, we show how to determine $\anchor = x$ in $\orderof{\ell}$ time. As described above, $\str[j + h..j + \ell)$ is a Lyndon run of at least four repetitions of a Lyndon word $\rungreek$. Consequently, $\str[j + \floor{\ell/4}..j + \ell)$ has the form $\text{suf}(\rungreek)\cdot\rungreek^t\cdot\text{pre}(\rungreek)$ with $t \geq 2$, where $\text{suf}(\rungreek)$ and $\text{pre}(\rungreek)$ are a proper suffix and a proper prefix of $\rungreek$. A string of this form is called \emph{extended Lyndon run}. Determining whether or not $\str[j + \floor{\ell/4}..j + \ell)$ is an extended Lyndon run, and also finding the period $\absolute{\rungreek}$ as well as the starting position $\absolute{\text{suf}}$ of the first full repetition, takes $\orderof{\ell}$ time and $\orderof{1}$ words of additional memory. This can be achieved using a trivial modification of Duval's algorithm for the Lyndon standard factorization \citep[Algorithm 2.1]{Duval1983}. For completeness, we describe it in \cref{cha:duvalextended}. 

If $\str[j + \floor{\ell/4}..j + \ell)$ is not an extended Lyndon run, then we have $\pssbps[o_j..o_{j + \floor{\ell/4} - 1}] = \pssbps[o_i..o_{i + \floor{\ell/4} - 1}]$ and no further steps are needed to satisfy \cref{lemma:ALmain}. Otherwise, we try to extend the extended Lyndon run to the left: We are now not only considering $\str[j + \floor{\ell/4}..j + \ell)$, but $\str[j..j + \ell)$. We want to find the leftmost index $j + h$ that is the starting position of a repetition of $\rungreek$. Given $\absolute{\rungreek}$ and $\absolute{\text{suf}({\rungreek})}$, this can be done naively by scanning $\str[j..j + \floor{\ell / 4}]$ from right to left, which takes $\orderof{\ell}$ time. If we have $h \geq \floor{\ell / 4} - \absolute{\rungreek}$, then the first case of \cref{lemma:ALmain} applies and no further steps are necessary. Otherwise, we let $\anchor \gets h + \absolute{\rungreek}$.
\newcommand{\curfbox}[1]{\strbox{\quad\ \clap{$\smash{#1}$}\quad\ }}%
\newcommand{\curlfbox}[1]{\curfbox{#1}}%
%
This concludes the proof of \cref{lemma:ALmain} and the description of our construction algorithm.

\section{Algorithmic Summary \& Adaptation to the Lyndon Array}

We now summarize our construction algorithm for the PSS tree. We process the indices from left to right using the techniques from \cref{sec:proci}, where processing an index means attaching it to the PSS tree. Whenever the critical time of processing an index is $\orderof{\ell}$, we skip the next $\Omega(\ell)$ indices by using the run extension (\cref{sec:runextend}) or the amortized look-ahead (\cref{sec:amolook}). Thus, the critical time per index is constant, and the total worst-case execution time is $\orderof{n}$. In terms of working space, we only need $\orderof{n \lg\lg n / \lg n}$ bits to support the operations described in \cref{sec:auxmaintain}. The correctness of the algorithm follows from the description. We have shown:

\begin{theorem}
	For a string $\str$ of length $n$ we can compute its succinct Lyndon array $\pssbps$ in $\orderof{n}$ time using $\orderof{n \lg\lg n / \lg n}$ bits of working space apart from the space needed for $\str$ and $\pssbps$.
\end{theorem}

\newcommand{\arr}[1]{{\mathcal{A}%
\def\temp{#1}\ifx\temp\empty
\else
  \xssarg{#1}%
\fi}}

The algorithm can easily be adapted to compute the Lyndon array instead of the PSS tree. 
For this purpose, we use a single array $\arr{}$ (which later becomes the Lyndon array), and no further auxiliary data structures. 
We maintain the following invariant:
At the time we start processing index $i$, we have $\arr{j} = \pss{j}$ for $j \in \pset_{i - 1}$, and $\arr{j} = \lyndarr[j]$ for $j \in [1, i) \setminus \pset_{i - 1}$. 
As before, we determine $p_m = \pss{i}$ with the techniques from \cref{sec:proci}. 
In Step 1 and Step 2 we require some access on elements of $\pset_{i - 1}$, which we can directly retrieve from $\arr{}$.
Apart from that, the algorithm remains unchanged.
Once we computed $p_m$, we set $\arr{i} \gets p_m$ ($=\pss{i}$). Additionally, it follows that $i$ is the first node that is not a descendant of any of the nodes $p_1, \dots, p_{m - 1}$, which means that we have $\nss{p_x} = i$ for any such node. Therefore, we assign $\arr{p_x} \gets i - p_x$ ($= \lyndarr[p_x]$).

The run extension and the amortized look-ahead remain essentially unchanged, with the only difference being that we copy and append respective array intervals instead of BPS substrings (some trivial shifts on copied values are necessary). Once we have processed index $n$, we have $\arr{j} = \pss{j}$ for $j \in \pset_{n}$, and $\arr{j} = \lyndarr[j]$ for $j \in [1, n] \setminus \pset_{n}$. Clearly, all indices $p_x \in \pset_{n}$ do not have a next smaller suffix, and we set $\arr{p_x} \gets n - p_x + 1 = \lyndarr[p_x]$. After this, we have $\arr{} = \lyndarr$.

Since at all times we only use $\arr{}$ and no auxiliary data structures, the additional working space needed (apart from input and output) is constant. The linear execution time and correctness of the algorithm follow from the description. Thus we have shown:

\begin{theorem}
	Given a string $\str$ of length $n$, we can compute its Lyndon array $\lyndarr$ in $\orderof{n}$ time using $\orderof{1}$ words of working space apart from the space needed for $\str$ and $\lyndarr$.
\end{theorem}

\FloatBarrier
\section{Experimental Results}

\newcommand{\evalfont}[1]{\mbox{\textsf{#1}}}
\newcommand{\laplain}{\evalfont{LA-Plain}}
\newcommand{\lasuc}{\evalfont{LA-Succ}}
\newcommand{\divsufsort}{\evalfont{DivSufSort}}
\newcommand{\lynisa}{\evalfont{LA-ISA-NSV}}

We implemented our construction algorithm for both the succinct and the plain Lyndon array (\lasuc{} and \laplain{}). The C++ implementation is publicly available at GitHub\footnote{\href{https://github.com/jonas-ellert/nearest-smaller-suffixes}{\texttt{https://github.com/jonas-ellert/nearest-smaller-suffixes}}}. As a baseline we compared the throughput of our algorithms with the throughput of \divsufsort\footnote{\href{https://github.com/y-256/libdivsufsort}{\texttt{https://github.com/y-256/libdivsufsort}}}, which is known to be the fastest suffix array construction algorithm in practice \citep{Fischer2017}. Thus, it can be seen as a natural lower bound for any Lyndon array construction algorithm that depends on the suffix array. Additionally we consider \lynisa{}, which builds the Lyndon array by computing next smaller values on the inverse suffix array (see \citep{Franek2016}, we use \divsufsort{} to construct the suffix array). For \lasuc{} we only construct the succinct Lyndon array without the support data structure for fast queries. All experiments were conducted on the LiDO3 cluster\footnote{\href{https://www.lido.tu-dortmund.de/cms/de/LiDO3/index.html}{\texttt{https://www.lido.tu-dortmund.de/cms/de/LiDO3/index.html}}}, using an Intel Xeon E5-2640v4 processor and 64GiB of memory. We repeated each experiment five times and use the median as the final result. All texts are taken from the Pizza \& Chili text corpus\footnote{\href{http://pizzachili.dcc.uchile.cl/}{\texttt{http://pizzachili.dcc.uchile.cl/}}}.

\begin{table}[t]
\small
\newcommand{\textlabel}[1]{\raisebox{.25em}{\rotatebox{40}{\rlap{\texttt{#1}}}}\phantom{0.00}}
\begin{tabular}{c|rrrrrr|rrrr}
	\rule{0cm}{1.8cm} & \textlabel{english.1GiB} & \textlabel{dna} & \textlabel{pitches} & \textlabel{proteins} & \textlabel{sources} & \textlabel{xml} & \textlabel{cere} & \textlabel{einstein.de} & \textlabel{fib41} & \textlabel{kernel}\\
	\hline\vphantom{${I}^{\strut}$}
	\laplain    & 60.57 & 50.83 & 60.58 & 62.18 & 66.13 & 82.10 & 53.08 & 59.09 & 41.71 & 62.27 \\
	\lasuc      & 52.81 & 46.03 & 49.49 & 52.77 & 57.31 & 68.56 & 48.20 & 50.35 & 35.30 & 54.42 \\
	\lynisa     &  4.61 &  4.86 &  9.13 &  4.40 &  7.41 &  7.11 &  5.44 &  6.72 &  3.81 &  6.79 \\
	\divsufsort &  5.53 &  5.76 & 11.61 &  5.21 &  9.25 &  8.62 &  6.57 &  8.45 &  4.20 &  8.45
\end{tabular}
\caption{Throughput in MiB/s.}
\label{tab:throughput}
\end{table}

\cref{tab:throughput} shows the throughput of the different algorithms. We are able to construct the plain Lyndon array at a speed of between 41 MiB/s (\texttt{fib41}) and 82 MiB/s (\texttt{xml}), which is on average 9.9 times faster than \lynisa{}, and 8.1 times faster than \divsufsort{}. Even in the worst case, \laplain{} is still 6.8 times faster than \lynisa{}, and 5.2 times faster than \divsufsort{} (\texttt{pitches}). When constructing the succinct Lyndon array we achieve around 86\% of the throughput of \laplain{} on average, but never less than 81\% (\texttt{pitches}).

In terms of memory usage, we measured the additional working space needed apart from the space for the text and the (succinct) Lyndon array. Both \laplain{} and \lasuc{} never needed more than 0.002 bytes of additional memory per input character (or 770 KiB of additional memory in total), which is why we do not list the results in detail.

\section{Conclusions \& Future Work}

We gave a more intuitive interpretation of the succinct Lyndon array, and showed how to construct it in linear time using $\orderof{n \lg\lg n / \lg n}$ bits of working space. The construction algorithm can also produce the (non-succinct) Lyndon array in linear time using only $\orderof{1}$ words of working space. There exist no other linear time algorithms that achieve these bounds.

Since our algorithm already processes the input text from left to right, it would be interesting to see if it can be adapted to become an online algorithm. Also, considering that next smaller \emph{values} can be efficiently computed in parallel \citep{Berkman1993}, there might be efficient parallel algorithms for the Lyndon array as well.
We also envision applications of our practical algorithms in full-text indexing, such as an improved implementation of Baier's suffix array construction algorithm \citep{Baier2016}, or as a first step in sparse suffix sorting \citep{fischer16deterministic,bille16sparse}.

\cleardoublepage
\bibliographystyle{abbrvnat}
\bibliography{biblio}

\cleardoublepage
\appendix

\section{Detecting Extended Lyndon Runs}

\label{cha:duvalextended}

In this section, we explain how to efficiently detect if a string is an extended Lyndon run. Recall that such a string has the form $\str = {\text{suf}(\rungreek)\cdot\rungreek^t\cdot\text{pre}(\rungreek)}$ with $t \geq 2$, where $\rungreek$ is a Lyndon word, and $\text{suf}(\rungreek)$ and $\text{pre}(\rungreek)$ are a proper suffix and prefix of $\rungreek$ respectively. 
\begin{alignat*}{1}
\str = \curfbox{\text{suf}(\rungreek)}{\underbrace{\curlfbox{\rungreek}\curlfbox{\rungreek}\curlfbox{\cdots}\curlfbox{\rungreek}\curlfbox{\rungreek}}_{\displaystyle t\text{ times}}}\curfbox{\text{pre}(\rungreek)}
\end{alignat*}

We use a simple modification of \citep[Algorithm 2.1]{Duval1983} to realize the detection mechanism. Originally, Duval's algorithm computes the uniquely defined \emph{Lyndon factorization} (sometimes called \emph{Lyndon decomposition}) of a string:

\begin{lemma}[\citealp{Chen1958}]\label{lemma:lynfac}
    Let $\str$ be a non-empty string. There exists a decomposition of $\str$ into non-empty factors $s_1, s_2,\dots, s_m$ such that all of the following conditions hold:
    \begin{enumerate}
        \vspace{-.75em}
        \item $\str = s_1\cdot s_2\cdot{\dots}\cdot s_m$
        \vspace{-.5em}
        \item $\forall i \in [1, m] : s_i \text{ is a Lyndon word}$
        \vspace{-.5em}
        \item $\forall i \in [2, m] : s_{i - 1} \geqlex s_i$
    \end{enumerate}
    There is exactly one such factorization for each string.
\end{lemma}

Now we show that the longest factor in the Lyndon factorization of an extended Lyndon run is exactly the repeating Lyndon word $\rungreek$ of the run. This makes it easy to detect if a string is an extended Lyndon run: We can simply compute the Lyndon factorization and determine the length of the longest factor. After that, a trivial postprocessing is sufficient to determine if the string actually is an extended Lyndon run.

\Needspace{6\baselineskip}\begin{lemma}\label{lemma:extlynrunfac}
    Let $\str = {\textnormal{suf}(\rungreek)\cdot\rungreek^t\cdot\textnormal{pre}(\rungreek)}$ be an extended Lyndon run. Let $x_1, \dots, x_{k_1}$ be the Lyndon factorization of $\textnormal{suf}(\rungreek)$, and let $y_1, \dots, y_{k_2}$ be the Lyndon factorization of $\textnormal{pre}(\rungreek)$. Then the Lyndon factorization of $\str$ is given by:
    \begin{alignat*}{1}
    \str = x_1 \cdot {\ \dots\ } \cdot x_{k_1}\ \cdot\ \ \underbrace{\rungreek \cdot \rungreek \cdot {\ \dots\ } \cdot \rungreek \cdot \rungreek}_{\displaystyle\textnormal{$t$ times}}\ \  \cdot\ \ y_1 \cdot {\ \dots\ } \cdot y_{k_2}
    \end{alignat*}\vspace{-1\baselineskip}
    \begin{proof}
        Clearly, the first two conditions of \cref{lemma:lynfac} are satisfied. We only have to prove the third one. Since we defined $x_1, \dots, x_{k_1}$ and $y_1, \dots, y_{k_2}$ to be the Lyndon factorizations of $\text{suf}(\rungreek)$ and $\text{pre}(\rungreek)$ respectively, we already know that $\forall i \in [2, k_1] : x_{i - 1} \geqlex x_i$ and $\forall i \in [2, k_2] : y_{i - 1} \geqlex y_i$ hold. Also, we trivially have $\rungreek \geq \rungreek$. Therefore, in order to prove that the third condition of \cref{lemma:lynfac} is satisfied, we only have to show that $x_{k_1} \geqlex \rungreek \geqlex y_1$ holds. Since $x_{k_1}$ is a non-empty suffix of $\text{suf}(\rungreek)$ and thus also a non-empty proper suffix of $\rungreek$, it follows that $x_{k_1} \glex \rungreek$ holds. Since $y_1$ is a prefix of $\text{pre}(\rungreek)$ and thus also a prefix of $\rungreek$, it follows (by definition of the lexicographical order) that $\rungreek \glex y_1$ holds. Therefore, the third property of \cref{lemma:lynfac} is satisfied.
    \end{proof}
\end{lemma}

If we look at the factors $x_i$ and $y_i$ of the factorization in the lemma, then each one of them is shorter than $\rungreek$, which means that $\rungreek$ is the longest factor of the factorization. Next, we explain how to exploit this property to detect if a string is an extended Lyndon run.

\subsection{Algorithmic Approach}

We start by taking a closer look at Duval's algorithm. Let $\str$ be a string with Lyndon factorization $s_1,\dots, s_m$. In Duval's original version of the algorithm, each factor is represented by its end position, i.e.\ the algorithm outputs a list $d_1, \dots, d_{m}$ of indices with $\forall i \in [1, m] : d_i = \sum_{j = 1}^{i} \absolute{s_j}$. 
Our algorithm for the detection of extended Lyndon runs uses this list as a prerequisite. Pseudocode is provided in \cref{algo:extlyndetec}.

Given any string that is not a Lyndon run, the algorithm outputs $\bot$. If however an extended Lyndon run is given, the algorithm outputs the period $\absolute{\rungreek}$ of the run, as well as the starting position $\absolute{\text{suf}(\rungreek)} + 1$ of the first full repetition of $\rungreek$. The algorithmic approach is simple: First, we only try to find the longest factor and its starting position. This can be achieved by processing the indices $d_1,\dots,d_m$ from left to right. Clearly, the length of factor $s_i$ is exactly $d_i - d_{i - 1}$ (if we define $d_0 = 0$). The starting position of factor $s_i$ is $d_{i - 1} + 1$. Initially, the longest known factor is $s_1$ with length $l = d_1$ and starting position $z = 1$ (lines 2--3). Then, we look at one factor at a time (line 4) and update the values of $l$ and $z$, whenever we find a factor that is longer than all previous ones (lines 5--7). After the last iteration of the loop, we know the length and starting position of the longest factor. Note that if the given string actually is an extended Lyndon run, then the starting position $z$ belongs to the \emph{first} occurrence of $\rungreek$. This holds because we process the factors in left-to-right order, and we specifically do \emph{not} update $z$ when finding a factor of equal length. 

Next, we have to verify if the computed values of $l$ and $z$ belong to an extended Lyndon run. Since such a run must have at least two repetitions, the period cannot be larger than $\floor{n / 2}$. Therefore, we first check if $2l > n$ holds, and return $\bot$ if that is the case (lines 8--9). Otherwise, we perform a single scan over $\str$ and check for each character if it equals the character that is located $l$ positions before (lines 10--11). If we find a mismatch, then we return $\bot$ (line 12). Otherwise, the string is an extended Lyndon run and we return $l$ and $z$ (line 13).


\begin{algorithm}
    \begin{algorithmic}[1]
        \Require A string $\str$ of length $n$
        \Ensure If $\str$ is an extended Lyndon run: Period $\absolute{\rungreek}$ and suffix length $\absolute{\text{suf}(\rungreek)}$.
        \color{white}\Ensure\color{black} If $\str$ is not an extended Lyndon run: $\bot$.
        \vspace{.5em}
        \State $d_1, \dots, d_m \gets \text{end positions of all factors of } \str$
        \State $l \gets d_1$
        \State $z \gets 1$
        \For{$i \in [2, m]$ in ascending order}
            \If{$d_i - d_{i - 1} > l$}
                \State $l \gets d_i - d_{i - 1}$
                \State $z \gets d_{i - 1} + 1$
            \EndIf
        \EndFor
        \vspace{.5em}
        \If{$2l > n$}
        \State \Return $\bot$
        \EndIf 
        \vspace{.5em}
        \For{$i \in [l + 1, n]$}
        \If{$\str[i - l] \neq \str[i]$}
        \State \Return $\bot$
        \EndIf
        \EndFor
        \vspace{.5em}
        \State \Return $l, z$
    \end{algorithmic}
    \caption{Detection of Extended Lyndon Runs}
    \label{algo:extlyndetec}
\end{algorithm}

\begin{lemma}
    \cref{algo:extlyndetec} detects if a string of length $n$ is an extended Lyndon run in $\orderof{n}$ time using $\orderof{1}$ words of memory apart from input and output.
    \begin{proof}
        The correctness follows from \cref{lemma:extlynrunfac} and the description above. We only have to prove the time and space bounds. In terms of execution time, we use Duval's algorithm to compute the indices $d_1, \dots, d_m$, which takes $\orderof{n}$ time \citep[Algorithm 2.1, Theorem 2.1]{Duval1983}. This clearly dominates the execution time of \cref{algo:extlyndetec}. It remains to be shown that $\orderof{1}$ words of memory are sufficient. Duval's algorithm computes the indices $d_1,\dots,d_m$ in a greedy manner, i.e.\ it outputs the indices one at a time and in left-to-right order. Since \cref{algo:extlyndetec} also processes the indices in left-to-right order, it is never necessary to keep more than two indices in memory at the same time. \Needspace{5\baselineskip}Therefore, we can interleave the execution of Duval's algorithm and \cref{algo:extlyndetec} such that we only compute the next index $d_i$ once it is actually needed. Apart from input and ouput, Duval's algorithm uses $\orderof{1}$ words of memory \citep[Theorem 2.1]{Duval1983}. Since \cref{algo:extlyndetec} only needs to keep the variables $l$, $z$, and two indices $d_i$ and $d_{i - 1}$ in memory, the additional memory usage is bound by $\orderof{1}$ words.
    \end{proof}
\end{lemma}

\end{document}